\documentclass[aip,pof,groupedaddress,reprint,onecolumn]{revtex4-1}

\usepackage{hyperref}

\usepackage{amsfonts}
\usepackage{amsmath}
\usepackage{amssymb} 

\usepackage{mathtools} 

\usepackage{multirow} 

\usepackage{siunitx}
\usepackage[caption=false]{subfig} 

\usepackage{graphicx}
\graphicspath{{figures/images/}{figures/plots/}}

\newcommand{\refe}{\operatorname{ref}}
\newcommand{\op}{\operatorname}
\newcommand{\deriv}[2]{\frac{\mathrm{d} #1}{\mathrm{d} #2}}
\newcommand{\partderiv}[2]{\frac{\partial #1}{\partial #2}}

\begin{document}

\title{Detonation propagation in annular arcs of condensed phase
  explosives}

\author{Eleftherios Ioannou}
\email[Corresponding author: ]{ei233@cam.ac.uk}

\author{Stefan Schoch}

\author{Nikolaos Nikiforakis}

\author{Louisa Michael}

\affiliation{Laboratory for Scientific Computing, Cavendish
  Laboratory, Department of Physics, University of Cambridge, CB3
  0HE, UK}

\date{\today}
 
\begin{abstract}
  We present a numerical study of detonation propagation in unconfined
  explosive charges shaped as an annular arc (rib). Steady detonation
  in a straight charge propagates at constant speed but when it enters
  an annular section, it goes through a transition phase and
  eventually reaches a new steady state of constant angular
  velocity. This study examines the speed of the detonation wave along
  the annular charge during the transition phase and at steady state,
  as well as its dependence on the dimensions of the annulus. The
  system is modeled using a recently proposed diffuse-interface
  formulation which allows for the representation of a two-phase
  explosive and of an additional inert material. The explosive
  considered is the polymer-bonded TATB-based LX-17 and is modeled
  using two JWL equations of state and the Ignition and Growth
  reaction rate law. Results show that steady state speeds are in good
  agreement with experiment. In the transition phase, the evolution of
  outer detonation speed deviates from the exponential bounded growth
  function suggested by previous studies. We propose a new description
  of the transition phase which consists of two regimes. The first is
  caused by local effects at the outer edge of the annulus and leads
  to a dependence of outer detonation speed on angular position along
  the arc. The second regime is induced by effects originating from
  the inner edge of the annular charge and leads to the deceleration
  of the outer detonation until steady state is reached. The study
  concludes with a parametric study where the dependence of the steady
  state and the transition phase on the dimensions of the annulus is
  investigated.
\end{abstract}



\maketitle 

\section{Introduction}

Arbitrarily-shaped explosive charges are used in a variety of modern
applications, particularly in the mining industry. The propagation of
detonations in such charges is influenced by the geometry of the
charge and exhibits different behavior from the case of straight
charges.
This study uses numerical simulations to investigate detonation propagation in
unconfined annular explosive charges in air. The aim is
to identify the characteristic features of detonation in such geometries and to
determine its dependence on the dimensions of the annular charge.

\begin{figure} 
  \centering 
  \begin{minipage}{0.38\linewidth}
    \includegraphics[width=\linewidth]{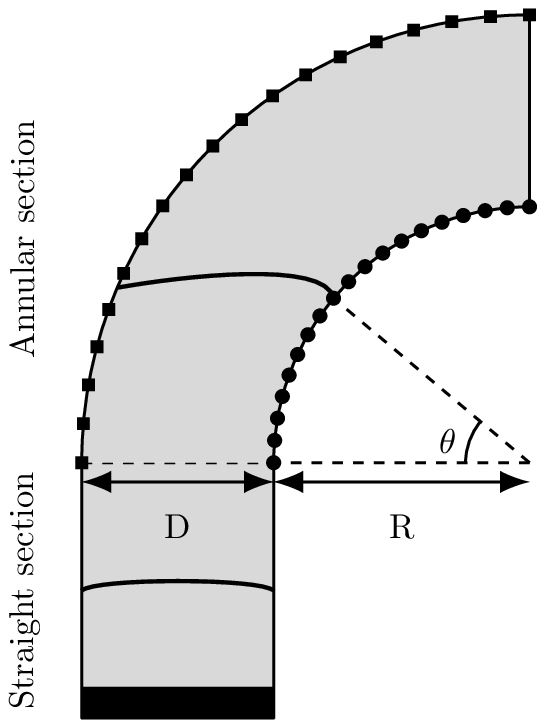}
    \caption{Illustration of the annular charge configuration. The
      annulus is of width $D$ and inner radius $R$. The angular
      position is defined by angle $\theta$. The detonation is
      initiated by a booster shown as the dark region and the solid
      thick lines represent the detonation front. The square and
      circular points represent the electrical pins used to measure
      detonation arrival times in the Lyle experiment.}
    \label{fig:lyle_configuration}
  \end{minipage} \qquad
  \begin{minipage}{0.52\linewidth}
    \includegraphics[width=\linewidth]{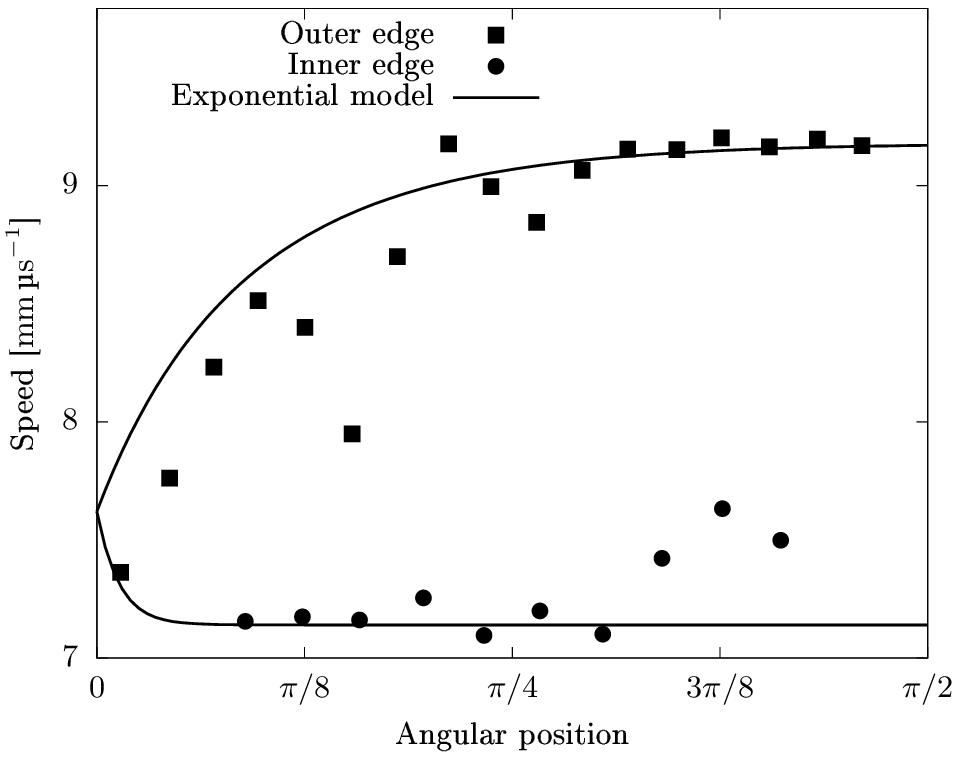}
    \caption{Experimental measurements of detonation speeds along the
      outer and inner boundaries of the explosive annulus adopted from
      the Lyle experiment~\cite{Souers1998}. The black curves
      correspond to the exponential model suggested by Souers et
      al.~\cite{Souers1998} to describe the evolution of the
      detonation speed along the annulus.}
    \label{fig:lyle_results}
  \end{minipage}
\end{figure}

This work is guided by two experimental studies on
annular charges. The first was performed by Lyle and Hayes at the
Lawrence Livermore National Laboratory (LLNL) in the 1980s and is
presented in a study by Souers et al.~\cite{Souers1998}. These used
unconfined charges of various compositions of LX-17 shaped as a
$90^\circ$ annular arc with a square cross section. The detonation was
initiated in a straight charge and was left to reach steady state
before entering the annular section, as illustrated in figure
\ref{fig:lyle_configuration}. A series of electrical pins were used at
the edges of the charge to measure arrival times of the detonation
wave, in addition to a streak camera used to capture the shape of the
detonation front. The second experiment was performed by Lubyatinsky et
al.~\cite{Lubyatinsky2004} who presented results of arrival times and
front break-out traces of detonation waves propagating in $180^\circ$
annular arcs of an unspecified high explosive. This experiment
involved explosive annuli of various widths
and radii confined by two types of material, namely steel and
PMMA.

Experiments on annular charges were also performed for the purpose of
calibrating and validating mathematical models for detonation propagation such
as Detonation Shock Dynamics (DSD). Tonghu et al.~\cite{Tonghu1998} studied
$60^\circ$, $90^\circ$ and $125^\circ$ annular arcs of a TATB-based
explosive. They obtained detonation arrival times and front shapes using arrays
of electrical pins and high speed photography and compared them against
numerical results from a DSD computational code. Similarly, Bdzil et
al.~\cite{Bdzil2003} measured the speed and front shape for a detonation exiting
a $135^\circ$ unconfined annular arc of PBX-9502. This was used for the
validation of time-dependent~\cite{Hill2010} and steady~\cite{Short2016}
DSD calculations and were found to be in good agreement with the experimental results.

Similar experiments have been performed for gaseous explosives motivated by
the development of rotating detonation engines. Nakayama et
al.~\cite{Nakayama2012} studied annular configurations of different
inner radii and same width for a range of characteristic detonation cell
width. They classified detonation propagation in different modes namely unstable,
critical and stable depending on the magnitude of the variation of inner normal
detonation velocity. A condition for stable propagation was determined based on the ratio of inner
radius to characteristic cell width. In addition it was found that a scaled
$D_n(k)$ relation exists which is almost independent of the configuration
parameters, inner radius and characteristic cell width.

A common outcome of the aforementioned studies is that detonation propagation in
annular charges settles to a steady state. This is characterized by constant
angular velocity of the detonation wave as opposed to constant linear velocity
observed in straight charges. Furthermore, there is a dependence of the steady
angular velocity on the charge dimensions, similar to the diameter effect seen
in straight charges. In particular, the reciprocal steady angular velocity has
an affine dependence on the inner radius of the annular
arc~\cite{Lubyatinsky2004}.

The shape of the detonation front in annular charges is asymmetric, in
contrast to straight charges which is symmetric about the
center line. Its leading peak is close to the inner edge as seen in
figure \ref{fig:lyle_configuration}. Its position depends on the
confining material which also affects the speed of the leading
peak. Materials with high impedance induce higher leading peak speeds
and smaller distances between the leading peak and the inner
edge~\cite{Lubyatinsky2004}.

The condition of constant angular velocity results in notably
different detonation speeds at each edge of the
front. Denoting the steady detonation speed measured along the inner
edge as $V_{\op{S}}$ and along the outer edge as $W_{\op{S}},$ the
constant angular velocity condition translates to
\begin{equation}
  W_{\op{S}} = \left( 1 + \frac{D}{R} \right) V_{\op{S}},
\end{equation}
where $R$ and $D$ are the inner radius and width of the arc
respectively, as depicted in figure \ref{fig:lyle_configuration}. It
is evident that depending on the ratio of width to radius, steady
state velocity measured along the outer edge of the charge can have a
significantly larger magnitude than at the inner edge. 

There is a fundamental difference in the nature of speeds measured along the
inner and outer edge of the annulus. At the inner edge, the detonation front
velocity is tangent to the boundary of the annular charge and the measured speed
is the actual speed at which the detonation propagates along that edge. However,
the detonation velocity at the outer edge is not tangent to the boundary of the
annulus and does not propagate along the outer boundary. Instead, the outer edge
of the detonation at any moment has originated from a previously interior part
of the detonation front. Thus, the measured speed is an apparent speed that the
detonation front exhibits along the curve defined by the outer boundary as if it
is moving across it. This explains why the speed measured at the outer boundary
can be significantly higher than the ideal CJ velocity, which would otherwise be
considered nonphysical for self-propagating detonations.

The transition phase is defined as the period beginning when the
detonation enters the annular section until steady state is
reached. The inner part of the detonation reaches steady state earlier
than the outer part which has a longer transition phase, as can be
seen from the results of the Lyle experiment in figure
\ref{fig:lyle_results}.

Although the dynamics of the detonation wave during the transition
phase have not been thoroughly explained, it is suggested to involve a process where equilibrium is achieved through an
energy flow across the detonation front~\cite{Souers2002}. Souers et
al.~\cite{Souers1998} suggest that the evolution of the inner and
outer speeds during the transition to steady state follows an
exponential function of time given by
\begin{equation} 
  u(t) = U_{\op{S}} + ( u_{\op{S}} - U_{\op{S}})\left(
1-\exp\left(\frac{-t}{\tau}\right)\right),
\end{equation} 
where $u$ is the wave speed at either edge, $u_{\op{S}}$ is the corresponding steady
speed, $U_{\op{S}}$ is the steady speed in
the straight section and $\tau$ is a time constant. The time constant is different
for the inner and outer edge and it determines the extent of the
transition period. It is derived by
approximating the speed of sound as three quarters of the steady
detonation speed in the straight section and is given by
\begin{equation} 
  \tau = \frac{4\Delta D}{3U_{\op{S}}},
  \label{eqn:eq_time}
\end{equation} 
where $\Delta D$ is the distance of the corresponding edge from the
leading peak of the detonation front. Plots of the exponential
functions that are suggested to describe the detonation
front speed at the inner and outer edge of the explosive charge are shown
in figure \ref{fig:lyle_results}.

Detonations in annular charges have also been investigated by a number of
numerical studies. Short et al.~\cite{Short2016} studied steady solutions of the
DSD model for a 2D annular charge using both numerical methods and asymptotic
analysis. The results showed a multi-layer structure of the detonation and
determined the dependence of angular speed and front structure on the size of
the annulus and on different degrees of confinement. The analysis distinguishes
between thin and thick arcs where different approximations are made for the
ratio of width to inner radius. For thick arcs, the steady angular speed
corresponds to the Huygens limit with correction terms which depend only on the
inner confinement.

Souers et al.~\cite{Souers1998} compared results
from the Lyle and Hayes experiments against numerical simulations
using the LLNL production code VHEMP with program burn and the
LASL-DYNA2D hydrodynamic code with the Ignition and Growth reaction
rate model. The simulations were performed with a resolution of
$\Delta x=\SI{500}{\micro\metre}$ at most and show good agreement with
the experimentally measured times to steady state. However, detonation
speed is overestimated by VHEMP and underestimated by DYNA2D and
detonation fronts have larger curvatures compared to the experimental
results.

Similarly, V\'agenknecht and Adam\'ik~\cite{Vagenknecht2006} performed
three-dimensional numerical simulations of the same experiments using LS-DYNA
and the beta burn model. They used a resolution of $\Delta
x=\SI{500}{\micro\metre}$ and reported good agreement with the experimental
detonation front curvature parameters but they also underpredicted the steady
speed values compared to the experiment. Tarver and Chidester~\cite{Tarver2007}
used the Ignition and Growth model and a resolution of $\Delta
x=\SI{50}{\micro\metre}$ to simulate several of the aforementioned experiments
on detonations in annular charges. The focus was on the steady state speeds of
the detonation front for which they showed good agreement with experimental
data.

The work presented here extents on the outcomes of previous studies on
detonations in annular charges. It uses direct numerical simulations to
present a complete description of the propagation of detonation along
the annular arc. Particular focus is given to the transition phase and
the identification of the effects that govern the evolution of the detonation
wave during this phase.

The numerical solution is used to calculate the detonation speed along
the inner and outer edge of the annular charge with respect to angular
position and to time. The steady state speeds show good agreement with
experimental results. However, the evolution of outer detonation speed
during the transition phase deviates from the suggested exponential
model. We propose a new description of the transition phase and show
that it can be divided into two regimes. In the first regime, the
outer detonation speed is governed by local effects at the outer
boundary which lead to a dependence of detonation speed on angular
position. In the second regime, effects originating from the inner
boundary reach the outer edge and bring the detonation to steady
state.

This work concludes with a parametric study where the inner radius and
width of the annular charge are varied. This reveals the dependence of
the transition phase and the steady state on the dimensions of the
annular arc. We show that the reciprocal steady angular velocity has
an affine relation with inner radius of the arc which was also
observed by Lubyatinsky et al.~\cite{Lubyatinsky2004}. The dependence
of the extent of the transition phase on the dimensions of the annular
charge is studied in terms of angle and time to steady state. Both
increase with width, whereas larger radii lead to a decrease of the
angle at which steady state is reached. In terms of time, there is
opposing behavior between configurations of small and large
widths. The transition duration increases with radius for large widths
but decreases for smaller widths.

The system is modeled using the recently proposed diffuse-interface method of Michael and
Nikiforakis~\cite{Michael2016}. The explosive considered is LX-17
(92.5\% TATB, 7.5\% Kel-F) which makes the results directly comparable
to the Lyle and Hayes experiments~\cite{Souers1998}. It is modeled
using two JWL equations of state and the Ignition and Growth reaction
rate law~\cite{Lee1980}. This choice was facilitated by the existence
of widely used sets of parameters for the particular explosive, as well
as by the availability of accessible experimental data that can be
used to validate the mathematical model and implementation of the
numerical methods.

The phenomenology of the Ignition and Growth model depends on the
resolution of the computations. Thus, particular care was given in
ensuring that the numerical simulations are adequately resolved. This
is facilitated by the use of a parallel, hierarchical,
block-structured adaptive mesh refinement framework.

This article continues with the presentation of the mathematical model
in section \ref{sect:model}. The governing equations as
well as the equations of state and reaction rate used to represent the
explosive are described. The employed numerical methods are discussed in section
\ref{sect:numerical_solution} and the implementation is validated in
section \ref{sect:validation}. The study of annular charges is
presented in section \ref{sect:annular} and a summary of the outcomes
and conclusions are discussed in section \ref{sect:conclusions}.

\section{Mathematical model}
\label{sect:model}

Numerical simulations involving explosives under confinement require a
mathematical formulation that can model the physical properties of
multiple materials and capture their interactions. Modeling
heterogeneous explosives, such as LX-17, poses a particular challenge
because their granular aggregate micro-structure is not possible to be
explicitly resolved in a mesoscale numerical simulation. Instead, it
is common to use a homogenized treatment which averages the fine scale
features of the explosive and accounts for the heterogeneous effects
by a phenomenological reaction rate law.

The mathematical model employed in this work was proposed by Michael
and Nikiforakis~\cite{Michael2016} and follows the approach described
above. It is a hybrid formulation for interfaces between immiscible
homogeneous fluids, where one of the materials is further divided into
two phases following the augmented Euler approach for modeling two
phase explosives. It assumes a continuum hydrodynamic representation
of the materials and allows for the modeling of an explosive with
distinct equations of state for the reactants and the products, and
also for an additional inert material.

This formulation is particularly suitable for modeling explosives
under compliant confinement because it can handle high density
gradients across interfaces, without the generation of spurious
oscillations in the solution. In addition, it allows for the use of
most types of equations of state and can be used for both ignition and
detonation propagation studies.

The mathematical model is defined by the following system of equations
\begin{equation}
\begin{aligned} 
\partderiv{z\rho_{\op{1}}}{t} + \mathbf{\nabla}\cdot z\rho_{\op{1}} \mathbf{u} &= 0, \\ 
\partderiv{(1-z)\rho_{\op{2}}}{t} + \nabla \cdot (1-z)\rho_{\op{2}} \mathbf{u} &= 0, \\ 
\partderiv{\rho \mathbf{u}}{t} + \nabla \cdot (\rho \mathbf{u}\otimes \mathbf{u} + pI) &= 0,
\\ \partderiv{\rho E}{t} + \nabla \cdot (\rho E+p) \mathbf{u} &= 0, \\ 
\partderiv{z}{t}+\mathbf{u}\nabla z &= 0, \\ 
\partderiv{z\rho_{\op{1}}\lambda}{t} + \mathbf{\nabla}\cdot z\rho_{\op{1}}\lambda \mathbf{u} &= z\rho_{1}\mathcal{R}.
\end{aligned}
\label{eqn:hybrid}
\end{equation} 
It features two continuity equations which represent the discrete
conservation of mass for each material as well as conservation laws
for the momentum and energy of the mixture. Quantities $\rho_1$ and
$\rho_2$ correspond to the density of the explosive and the inert
material respectively. Quantities $\rho$, $\mathbf{u}, p$ and $E$ are
the mixture density, velocity, pressure and total specific energy
defined as \[ E = e + \frac{\lVert \mathbf{u} \rVert }{2},\] where $e$
is the specific internal energy of the mixture.

The composition of the mixture is determined by the quantity
$z\in[0,1]$ which represents the volume fraction of the explosive and
is governed by an advection equation. Equivalently, quantity $1-z$ is
the volume fraction of the inert material. The explosive material is
further divided into two phases, which represent the reactants and the
products. We define $\lambda\in[0,1]$ as the mass fraction of the
reactants. This is also governed by an advection equation with a
source term $\mathcal{R}$ describing the chemical reactions that turn
reactants into products. However, it is combined with the continuity
equation of the explosive material and put into a conservative form
which represents the conservation of mass of the reactants. The
equations do not include any terms for viscous friction or heat
conduction as it is assumed that their effect is negligible in this
case study.

The formulation allows for the interface to diffuse on a small number
of computational cells over which a set of mixture rules has to be
defined. These rules relate the thermodynamic variables of the mixture
to those of the individual constituents. Considering mass and energy
as additive quantities, the mixture variables are given by
\begin{subequations}
\begin{align} 
\rho &= z\rho_{\op{1}} +
(1-z)\rho_{\op{2}}, \label{eqn:mixture_rules_1} \\ \rho e &= z\rho_{\op{1}}
e_{\op{1}} + (1-z)\rho_{\op{2}}
e_{\op{2}}, \label{eqn:mixture_rules_3}
\end{align}
\label{eqn:mixture_rules}
\end{subequations} where quantities with subscript 1 and 2 correspond
to the explosive and the inert material respectively. Subsequently, the
density and the specific internal energy of the two-phase explosive are
given by
\begin{subequations}
\begin{align} 
\frac{1}{\rho_{\op{1}}} &=
\lambda\frac{1}{\rho_{\op{a}}} +
(1-\lambda)\frac{1}{\rho_{\op{b}}}, \label{eqn:mixture_rules_2} \\
e_{\op{1}} &= \lambda e_{\op{a}} +
(1-\lambda)e_{\op{b}}, \label{eqn:mixture_rules_4}
\end{align}
\label{eqn:mixture_rules2}
\end{subequations} where subscripts $a$ and $b$ denote quantities of
the reactants and the products that comprise the explosive.

In addition to the mixture rules, the system requires closure
conditions to be fully determined. Between the explosive and the inert
material only one closure condition is necessary, as the density of
each material is readily available from the state variables of the
equations. Between the reactants and the products, two mixture rules
are required as only the total explosive density is known and root
finding procedures need to be followed to determine the individual
reactants and products densities. Here, the closure conditions chosen
are isobaric between the explosive and inert material as it has been
proven to give better stability at the interfaces~\cite{Allaire2002},
and isobaric and isothermal between the reactants and products as in
similar studies~\cite{Banks2007,Kapila2007,Tarver2005,Tarver2010}. For
more details on aspects of the mathematical model the reader is
referred to the work by Michael and Nikiforakis~\cite{Michael2016}.

\subsection{Equations of state}
\label{subsect:eos}

The definition of an equation of state (EOS) for each material is
required to close the system of equations of the mathematical
formulation. In this work, the LX-17 reactants and products are
modelled by two distinct Jones-Wilkins-Lee (JWL) equations of state
and the air is modeled as a perfect gas.

The JWL equation of state can be written in the Mie-Gr\"{u}neisen form
\begin{align} 
e - e_{\refe}(\rho) = \frac{p - p_{\refe}(\rho) }{\rho\Gamma(\rho)}, \\ 
T = \frac{e - e_{\refe}(\rho) }{c_{\op{v}}},
\label{eqn:mieGruneisen}
\end{align} with a constant Gr\"{u}neisen coefficient $\Gamma(\rho)=\Gamma_0$ and the
following reference curves
\begin{align} 
p_{\refe}(\rho) &= A\exp{\left(
-R_1\frac{\rho_{\op{0}}}{\rho}\right)} + B\exp{\left(
-R_2\frac{\rho_{\op{0}}}{\rho}\right)} ,\\ e_{\refe}(\rho) &=
\frac{A}{\rho_0 R_1}\exp{\left( -R_1\frac{\rho_0}{\rho}\right)} +
\frac{B}{\rho_0 R_2}\exp{\left( -R_2\frac{\rho_0}{\rho}\right)} - Q,
\end{align} 
where $A$, $B$, $R_1$ and $R_2$ are parameters calibrated for the particular explosive.

For the explosive products, the pressure reference curve of the JWL EOS
represents the isentrope through the Chapman-Jouguet (CJ) point and is
fitted to experimental data, usually from cylinder
tests~\cite{Fickett2000}. The energy reference curve is determined by
integrating the pressure reference curve, since
$\mathrm{d}e=-p\mathrm{d}v$ for an isentropic process. For the
reactants, the parameters are fitted to measurements of the Hugoniot
locus through an initial state.


Air is modeled by the perfect gas equation of state. This can
also be expressed in the Mie-Gr\"{u}neisen form through the trivial
reference curves
\begin{align}
p_{\refe} &= 0, \\
e_{\refe} &= 0,
\end{align}
and Gr\"{u}neisen coefficient $\Gamma=\gamma -1.$

\subsection{Reaction rate law}
\label{subsect:ignitiongrowth}

The reaction rate law used in this study is Ignition and Growth
(I\&G)~\cite{Lee1980} and plays an important role in 
capturing the reactive properties of the granular explosive within the homogeneous
representation of the material. The heterogeneity of the explosive material is
accounted for through this multi-stage, pressure-based reaction rate,
which provides a phenomenological description of the effects of the
micro-structure of the explosive using macroscopic material
parameters.

The I\&G reaction rate model is given by
\begin{equation} \mathcal{R} = \mathcal{R}_{\op{I}} +
\mathcal{R}_{\op{G_1}} + \mathcal{R}_{\op{G_2}},
\end{equation} and the three terms are defined as,
\begin{align} \mathcal{R}_{\op{I}} &= \begin{aligned}[t]
I(1-\phi)^b(\rho -1 -a)^x H(\rho/\rho_0 - 1 - \alpha)
H(\phi_{\op{ig}} - \phi ),
\end{aligned}
\label{eqn:ig_1}\\ \mathcal{R}_{\op{G_1}} &= G_1(1-\phi)^c\phi^d p^y
H(\phi_{\op{G_1}} - \phi), \label{eqn:ig_2}\\ \mathcal{R}_{\op{G_2}} &=
G_2(1-\phi)^e\phi^g p^z H( \phi - \phi_{\op{G_2}} ), \label{eqn:ig_3}
\end{align} where $H(x)$ is the Heaviside step function,
$\phi=1-\lambda$ is the mass fraction of the products and $\rho$ and
$p$ are the density and pressure of the explosive respectively. The rest of the
parameters are constants that are calibrated for each particular
explosive.

The I\&G model captures the complex ignition and burning
processes in heterogeneous explosives by using three terms to
represent the processes occurring in the
initiation and propagation of detonations in a heterogeneous
explosive~\cite{Kapila2007}. In the shock-to-detonation
transition process, ignition occurs due to shock-induced heating and friction
as well as hot-spot formation through cavity collapse in porous
explosives. These initiation mechanisms are represented in the ignition term
\eqref{eqn:ig_1} which is activated when density increases above a
threshold $\alpha$ and is used only in the initial stages of the
reaction.

The remaining two terms are called growth terms and have different
interpretations depending on whether the application involves
initiation or propagation of detonation. In the latter case, which
applies to this study, the growth terms represent the formation of the
products. In particular, term \eqref{eqn:ig_2} models the rapid
formation of gas products and term \eqref{eqn:ig_3} the slow
diffusion-controlled formation of solid carbon.

The JWL and I\&G parameters are selected and adjusted jointly to
accurately represent a specific explosive and application. For example
the parameters can be different between applications that involve
either ignition or propagation of detonation, even for the same
explosive.

\subsection{Data set and non-dimensionalization}

\begin{table}
  \centering
  \begin{tabular}{l@{\qquad}c@{\qquad}c}
    \hline
    \hline
    \multirow{2}{*}{Parameters} &  \multicolumn{2}{c}{LX-17} \\
    & Reactants & Products  \\ 
    \hline
    $\Gamma _0$ & 0.8938 & 0.5\\
    $A$ [\SI{e11}{\pascal}] & \num{778.1} & \num{14.8105} \\
    $B$ [\SI{e11}{\pascal}] & \num{-0.05031} & \num{0.6379}  \\
    $R_1$ & 11.3 & 6.2 \\
    $R_2$ & 1.13 & 2.2\\ 
    $c_V$  [\si{\metre\squared\per\second\squared\per\kelvin}] & \num{1305.5} & \num{524.9} \\
    $Q$ [\SI{e6}{\metre\squared\per\second\squared}] & 0 & \num{3.94}  \\
    $\rho _0$ [\si{\kilo\gram\per\metre\cubed}] & 1905 & 1905 \\
    \hline
    \hline
  \end{tabular}
  \caption{JWL EOS parameters for LX-17~\cite{Tarver2005}.}
  \label{tbl:eosparameters}
\end{table}

The parameter sets found in the literature for the equation of state and
reaction rate of a particular explosive can often vary. The variation
is attributed to the different experiments to which the
parameters were fitted but also to the particular process that the
parameter set intents to model. The parameter data set for the explosive LX-17
used in this study is taken from the work of Tarver~\cite{Tarver2005}
and has been used in similar
studies~\cite{Kapila2007,Tarver2010}. The parameters for
the JWL EOS are shown in table \ref{tbl:eosparameters} and the
reaction rate parameters are shown in table \ref{tbl:IGparameters}. This
set is suggested to be more suitable for detonation propagation rather
than initiation~\cite{Tarver2005}.

The equation of state and reaction rate parameter sets are often given
in a non-dimensional form, where particular parameters related to the
application were used as reference values in the
non-dimensionalization process.  Following the example of Kapila et
al.~\cite{Kapila2007} we use the CJ detonation speed of the explosive
as one of the reference values. This is calculated analytically using
the CJ theory~\cite{Lee2008}. The reference values used for
non-dimensionalization are
\begin{equation}
  \begin{aligned}
    \rho_0 = \rho_{\refe} &= \SI{1905}{\kilo\gram\per\meter\cubed} \\
    D_{\op{CJ}} = u_{\refe} &= \SI{7.6799}{\milli\meter\per\micro\second} \\
    t_{\refe} &= \SI{1}{\micro s}. \\
  \end{aligned}
  \label{eqn:reference_values_1}
\end{equation}
From the above, the reference values for the rest of the flow
variables can be calculated as
\begin{equation}
  \begin{aligned}
    \rho_0 D_{\op{CJ}}^2 = p_{\refe} &= \SI{112.359}{GPa} \\
    D_{\op{CJ}}^2 = e_{\refe} &= \SI{58.98}{\milli\meter\squared\per\micro\second\squared} \\
    D_{\op{CJ}} t_{\refe} = l_{\refe} &= \SI{7.6799}{mm}. \\
  \end{aligned}
  \label{eqn:reference_values_2}
\end{equation}

\begin{table*}
  \centering
\begin{ruledtabular}
  \begin{tabular}{l c l c l c}
    \multicolumn{2}{c}{$\mathcal{R}_{ig}$} &  \multicolumn{2}{c}{$\mathcal{R}_{G1}$} &  \multicolumn{2}{c}{$\mathcal{R}_{G2}$} \\
    \hline
    $I$ [\si{\per\second}] & \num{4.0e12} & $G_1[\si{\per\giga\pascal\cubed\per\milli\second}]$ & 4500 & $G_2[\si{\per\giga\pascal\per\milli\second}]$ & 30 \\

    $a$ & 0.22 & $b$ & 0.667 & $c$ & 0.667   \\

    $d$ & 1 & $e$ & 0.667 & $g$ & 0.667  \\

    $x$ & 7 & $y$ & 3 & $z$ & 1 \\

    $\phi_{ig}$ & 0.02 & $\phi_{G1}$ & 0.8 & $\phi_{G2}$ & 0.8 \\ 

  \end{tabular}
\end{ruledtabular}
  \caption{Ignition and Growth parameters for LX-17~\cite{Tarver2005}.}
  \label{tbl:IGparameters}
\end{table*}

\section{Numerical solution}
\label{sect:numerical_solution}

The mathematical formulation presented in the previous section
constitutes a non-linear hyperbolic system with source terms. Starting
from the initial conditions, the numerical solution is advanced in time
through the process of operator splitting. This allows for separate
solving of the homogeneous part using an appropriate hyperbolic solver
and the independent use of an ODE solver to compute the effect of the
source terms.

The hyperbolic part is solved using the finite volume method
MUSCL-Hancock~\cite{vanLeer1997}. It is a high-resolution, shock
capturing, Godunov-type reconstruction scheme which is second order
accurate in time and space. To avoid the spurious oscillations near
steep gradients in the flow that would otherwise occur in high order
schemes, we use the van Leer slope limiter~\cite{vanLeer1974} on the
primitive variables. The scheme requires a Riemann solver for
calculating fluxes at cell interfaces for which we use
HLLC~\cite{Toro1997}. The non-conservative equation of the volume
fraction is solved with the Godunov method for advection
equations~\cite{RichardSaurel2009}. The reaction rate source term, as
well as geometric source terms arising from axisymmetric problems in
cylindrical coordinates are solved using a 4th order Runge-Kutta
method.

The model and numerical solvers are implemented within a parallel,
hierarchical, block-structured adaptive mesh refinement (AMR)
computational framework~\cite{Schoch2015a} which provides increased
resolution at regions of interest. This allows for the computations to
be performed on a highly refined grid without a high computational
expense.

\subsection{Detonation front detection}

The work in this paper focuses on the analysis of the speed of a
detonation wave and therefore, it requires accurate detection of the
wave front and calculation of its speed. The position of the
detonation front in a simulation output can only be accurate to within
one grid cell at best. In addition, the use of a shock capturing
numerical method means that discontinuities are smeared over a few
grid cells which introduces additional uncertainty in the position of
the shock.

The algorithm used in this study for detecting shock position from the
numerical solution relies on the large pressure gradients across shock
waves compared to the rest of the domain. After every time-step, the
normalized gradient of pressure is calculated for every cell. The
position and state of cells that are above a predefined threshold
\begin{equation}
  \frac{\lVert\nabla p\rVert }{p} > 10,
\end{equation}
are output along with the simulation time. The extracted cells are
then grouped into cells that are at the same distance from the center
of the annulus within a range of the cell size $\Delta x$. The positions
of all cells in a group are then averaged and the calculated points mark
the detonation front position. The obtained data will unavoidably
have an error variance of a few computational cells but the above
method provides a consistent way of detecting front location which
does not dependent on how the discontinuity is smeared over the
cells.

The values of position over time are then used to calculate the
instantaneous speed of the detonation using a central differences
scheme. The calculation of derivatives from experimental or simulation
data greatly increases the noise levels in them. Thus, the stencil
used in this scheme is chosen to be wide, using up to ten data
points in each direction in order to provide a smooth representation
of the detonation front speed.

\begin{figure}
  \centering
  \includegraphics[width=\linewidth]{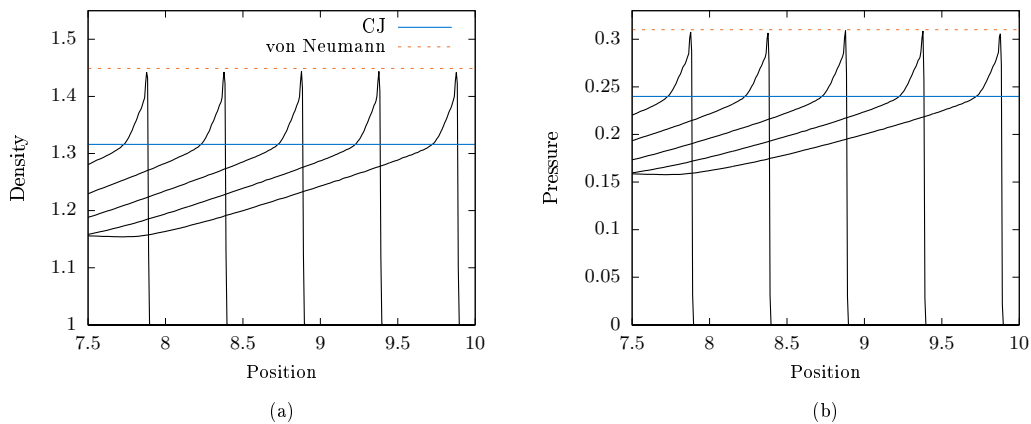}
  \caption{The numerical solution of the one-dimensional steady
    detonation in LX-17. The results are plotted at constant time
    intervals of $t=0.5$. The horizontal lines represent the
    analytically calculated values at the CJ and von Neumann points.}
  \label{fig:1Dznd}
\end{figure}

\section{Validation and grid convergence}
\label{sect:validation}

The mathematical model and implementation of the numerical methods are
assessed through a series of test problems to ensure their validity
and suitability for the considered application. These include the
study of the one-dimensional steady detonation and of the diameter
effect in cylindrical charges. In addition, a grid convergence study
is performed to establish the resolution for which the solution has
sufficiently converged.

\subsection{One-dimensional steady state detonation}
\label{subsect:1Dznd}

We consider the numerical solution of a one-dimensional steady
detonation. The structure of the detonation wave is described by the
ZND detonation model~\cite{Lee2008}. Characteristic quantities, such
as the states at the von Neumann and CJ points are calculated
analytically and are used to verify the implementation of the model.

The setup is one-dimensional and contains the explosive only. The
initial conditions consist of a small region of high pressure, equivalent to
a booster, placed at the left end of the domain and the rest of the
explosive is at ambient conditions. The pressure in the booster region
is set to 0.24 (\SI{27}{\giga\pascal}) which is close to the expected CJ
pressure of the explosive. This causes the rapid expansion of the
explosive in this region, which compresses and ignites the explosive
ahead, leading to the quick formation of a steady detonation wave.

The numerical solution of the one-dimensional detonation wave is shown
in figure \ref{fig:1Dznd} for a resolution of $\Delta
x=\SI{6.25e-3}{}$. The solution is presented in a series of density
and pressure plots, for times after the detonation has settled to
steady state. The von Neumann and CJ points of the numerical solution
match the values calculated analytically.

\subsection{Steady detonation in cylindrical charges}
\label{subsect:rate_sticks}

The two-dimensional implementation is validated using an unconfined
rate stick configuration. The setup is three-dimensional axisymmetric
and is solved in a two-dimensional domain with the addition of
geometric source terms. Each rate stick is defined by its radius $R$
and has a length of $L\approx10R$ which was found to be sufficient for
the detonation to settle to steady state before it reaches the end of
the charge. The detonation is initiated through a booster region of
high pressure, similar to the case of the 1D steady detonation.

\begin{figure}
  \includegraphics[width=\linewidth]{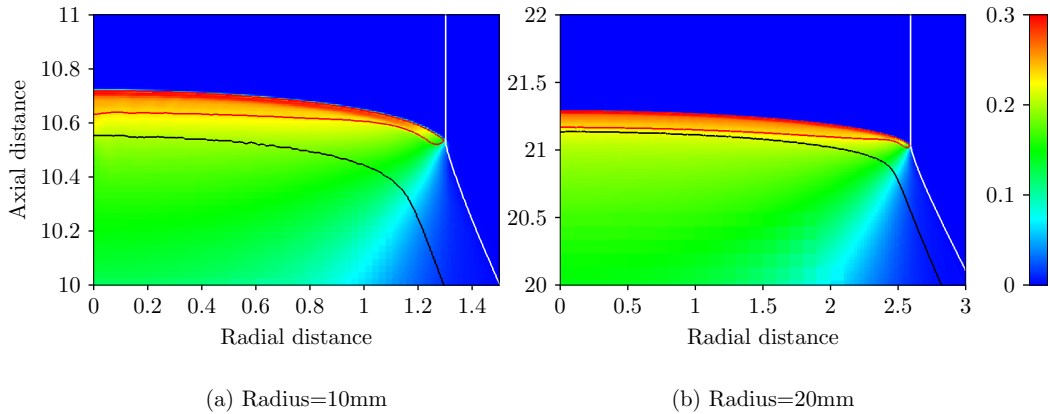} 
  \caption{Pressure plots of the structure of steady detonation
    waves propagating in cylindrical charges of different radii. The
    white line is the explosive-air interface, the black line
    corresponds to the reaction zone end (RZE) and the red line to
    the sonic locus which outlines the detonation driving zone
    (DDZ).}
  \label{fig:rate_stick_steady_state}
\end{figure}

A base grid resolution of $\Delta x=0.1$ is used, with two levels
of refinement, each with a refinement factor of 4, yielding an
effective resolution of $\Delta x=\SI{6.25e-3}{}$. It was ensured that
the solution has converged by performing a convergence study, as presented in section
\ref{subsect:convergence_study} for the case of annularly shaped
charges.

The numerical solutions of steady state detonation in rate sticks of radius
\SI{10}{mm} and \SI{20}{mm} are shown in figure
\ref{fig:rate_stick_steady_state}. As the detonation wave propagates,
the explosive products expand against the surrounding material which
results in a diverging flow as evidenced by the position of the
explosive-air interface. Also depicted is the reaction zone end
(RZE) locus, defined as the contour $\lambda = 0.001$ and the
sonic locus, defined as the curve which satisfies \[ M \equiv
\frac{\lVert\mathbf{v}\rVert}{c} =1, \] where $\mathbf{v}$ is the flow
velocity in the frame of the detonation and $c$ is the local speed of
sound.

The sonic locus outlines the region behind the shock in which the flow
is subsonic and is often referred to as the detonation driving zone (DDZ). Outside
the DDZ, the flow is supersonic and does not influence the propagation
of the detonation. In the steady detonations of figure
\ref{fig:rate_stick_steady_state}, the sonic locus intersects the
detonation front and not the explosive-air interface. This means that
the detonation wave is completely decoupled from the surrounding
material because the flow along the explosive-air interface is
supersonic. This classifies the configuration as unconfined, since the
surrounding material has no effect on the propagation of the
detonation.

The position of the sonic locus inside the reaction zone is
characteristic of the diverging flow in charges of finite radius. This
results in some of the chemical energy released in the reaction zone
not being supplied to the detonation front. In charges of
smaller radius, the curvature of the detonation front increases and the
sonic locus moves further away from the RZE which results in the DDZ
covering less of the reaction zone region. This translates to even
less energy being used to drive the detonation wave and gives rise to
the diameter effect in which the detonation slows down as the radius
of the charge decreases and eventually leads to detonation
failure. The difference in the position of the sonic locus with
respect to the RZE for rate sticks of different radii can be seen in
figure \ref{fig:rate_stick_steady_state}.

\begin{figure}
  \centering
    \includegraphics[scale=1]{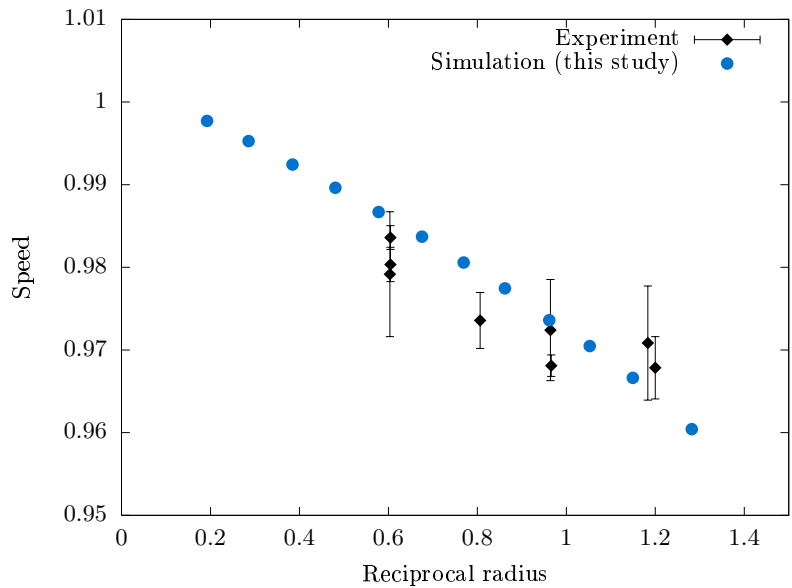}
  \caption{Numerical and experimental~\cite{Souers2009} steady detonation speeds over reciprocal radius for unconfined cylindrical LX-17 charges.}
  \label{fig:diameterEffect}
\end{figure} 

To assess the suitability of the mathematical model in capturing the
dependence of detonation speed on charge radius, we compare the
numerical results of detonation speeds against experimental
results. The experiments were performed at the Lawrence Livermore
National Lab~\cite{Souers2009} and used LX-17 charges confined by thin shells of either
copper or PMMA, as well as bare charges. They measured the average
detonation speed over the last third of the charge in which the
detonation is assumed to be in steady state. The PMMA shells had
thickness between \SIrange{1}{3.25}{\milli\meter} which is at most
25\% of the charge radius. These configurations are considered unconfined
and are used to compare against the results of the numerical
simulations.

The experimental and numerical detonation speeds can be seen in figure
\ref{fig:diameterEffect}. The diameter effect curve obtained from the
numerical solution is concave downwards which is typical of
heterogeneous condensed phase explosives~\cite{Sheffield2009}. The
numerical detonation speeds are within the range of values of the
experiment. The calculated failure radius is between
\SIrange{5.5}{6}{\milli\meter} which agrees with the numerical results
from Tarver and McGuire~\cite{Tarver2002a} and with their reported
experimental value of \SI{6}{mm}. The large error margins of the
experimental values do not allow for assessing the accuracy of the
numerical results. However, this test establishes the ability of the
model to capture the diameter effect for condensed phase high
explosives.

\subsection{Convergence study}
\label{subsect:convergence_study}

\begin{figure}
  \centering 
  \includegraphics[width=\linewidth]{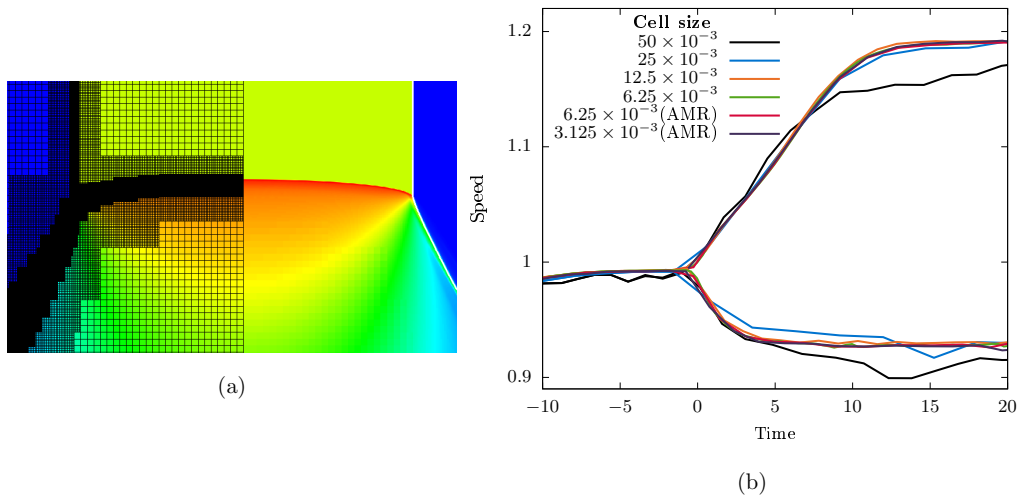}
  \caption{Convergence and AMR coverage study for the annular arc
    configuration. (a) Depiction of two levels of AMR covering the detonation
    front and material interface. (b) Detonation speed over time at the inner
    and outer boundary of the explosive, calculated from solutions of increasing
    resolution.}
  \label{fig:ResStudy}
\end{figure}

The use of sufficient resolution for the computations is essential in
capturing the phenomenology of the Ignition and Growth model. If the
reaction zone is not adequately resolved, it does not exhibit the
intended phenomenological description of the reaction processes in
the explosive and leads to critical loss of accuracy in the
solution. Bdzil et al.~\cite{Bdzil2003} studied a simplified pressure-based
reaction rate law and determined that predicting the detonation speed
of a straight cylindrical charge within \SI{10}{\metre\per\second} of
the actual value requires 50 or more cells in the reaction zone.

The convergence of the numerical solution is assessed by examining the
detonation speed over time for a series of simulations of increasing
resolution. We use the annular charge configuration and the detonation speed is
calculated along the inner and outer parts of the charge. The domain boundary
conditions are set to transmissive~\cite{Toro1997} to allow for waves to exit
the domain without any effect on the flow inside. This condition is not
perfectly satisfied in multi-dimensional problems and partial reflections occur
which influence the flow inside the domain. Thus, we ensured that the domain
boundaries are sufficiently far as to not influence the reaction zone and
interfere with detonation propagation.

The initial resolution is set to $\Delta x=0.05$ which corresponds to
\SI{384}{\micro\meter} and is halved for every subsequent simulation. In
addition, we ensure sufficient coverage of the important regions by the adaptive
mesh refinement (AMR) process. This is done by comparing the numerical solution
obtained using a uniform grid with one that used an AMR grid at the same
effective resolution. The refinement criterion is set using the density
gradient, i.e. \[\frac{\lVert\nabla \rho\rVert}{\rho} > 1.\]
This results in refining the detonation front and the reaction zone, as well as
the interface between the explosive and air due to the sharp density difference
between the two materials.

Figure \ref{fig:ResStudy} shows the speed of the detonation wave over
time at the inner and outer edge of the explosive charge for the set
of resolutions used. The solution gives indistinguishable detonation
speeds for resolutions higher than $\Delta x=\SI{6.25e-3}{} (\SI{48}{\micro m})$ which
is the resolution selected for this study. Moreover, identical
solutions are also obtained when utilizing AMR which ensures that the
refinement criterion results in sufficient coverage of the appropriate
regions.

\section{Detonation propagation in annular arc charges}
\label{sect:annular}

Having validated the mathematical model and the numerical algorithms
we now turn to the study of the propagation of detonations in
annularly shaped explosive charges. The configuration is as shown in
figure \ref{fig:lyle_configuration} and consists of an unconfined explosive
charge of rectangular cross section with a straight and an annular
section. The annular arc extents to $90^\circ$ and is defined by
the inner radius $R$ and width $D$. We assume that the explosive
charge is sufficiently long in the direction of the axis of curvature
to allow modeling the system as two-dimensional.

The explosive is ignited by the rapid expansion of a high pressure
region placed at the low end of the straight
charge. This leads to the quick formation of the detonation wave, which
reaches steady state in the straight section and subsequently enters
the annular section of the explosive charge. The detonation speed is
measured along the edges of the annulus and angular position is given
by angle $\theta$, also shown in figure \ref{fig:lyle_configuration}.

\subsection{The Lyle and Hayes experiments}

We initially study the configurations used in the Lyle and Hayes
experiments~\cite{Souers1998} for which experimental results are
available. These used unconfined LX-17 charges of different
dimensions. The Lyle experiment had an annular charge of inner radius
$R=\SI{88.9}{mm}$ and width $D=\SI{25.4}{mm}$, while the Hayes
configuration had $R=\SI{63.5}{mm}$ and $D=\SI{38.1}{mm}$. The
straight section was of length $L=\SI{116}{mm}$, which allowed enough
travel distance for the detonation to reach steady state before
entering the annular section.

\begin{figure}
  \centering
  \includegraphics[width=\linewidth]{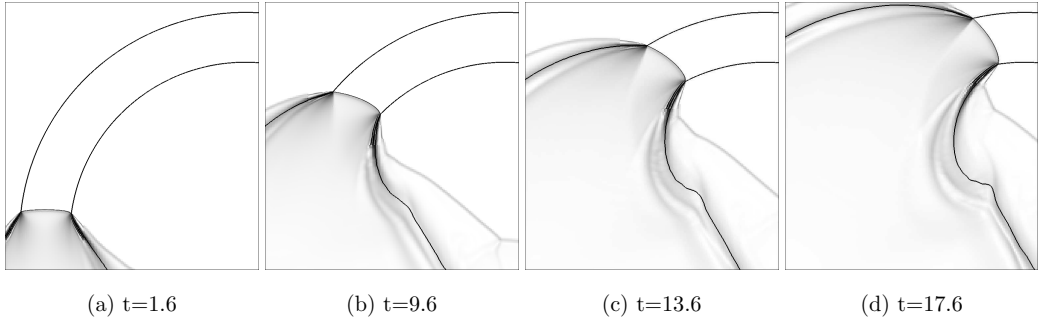}
  \caption{Density gradient plots of the propagation of a detonation wave in an
    annularly shaped explosive. The dark lines represent the interface between
    explosive and air as well as the detonation wave front. The shock developed in the
    air is shown with a lighter shade. The charge is of radius $R=11.578$
    and width $D=3.308$ which corresponds to the Lyle experiment.}
  \label{fig:bendy_plots}
\end{figure}

Figure \ref{fig:bendy_plots} shows density gradient plots of the detonation wave
propagating in the annular section of the explosive for the case of the Lyle
configuration. These plots accentuate the interface between the explosive and
air as well as the detonation wave front. The shock wave in air is shown with a
lighter shade. The results show the development of an apex in the explosive-air
interface. This is an effect of the different geometry between the straight and
annular sections. A steady detonation wave exerts a constant force on the
interface in the direction normal to the interface. In the straight section this
keeps the interface straight, whereas in the annular section each point of the
interface is pushed at different direction and the resulting interface is
curved. Thus, an apex develops at the point where the charge geometry changes.

\begin{figure}
  \centering
  \begin{minipage}[t]{0.46\linewidth}
    \includegraphics[width=\linewidth]{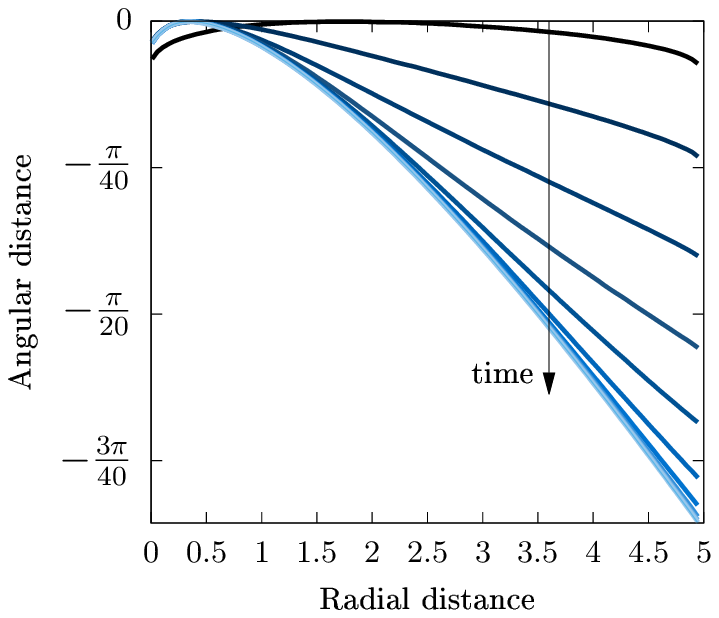}
    \caption{Evolution of the detonation front for the Hayes
      configuration. The fronts are shown at constant time intervals
      starting when the detonation enters the annular section (black)
      and move from dark to light color as time progresses. The
      curvature of the detonation front increases until it reaches
      steady state and maintains its shape. Radial distance is
      measured from the inner edge of the charge and angular distance
      from the leading peak of the front.}
    \label{fig:shock_fronts}
  \end{minipage} \qquad
  \begin{minipage}[t]{0.46\linewidth}
    \centering
    \includegraphics[width=\linewidth]{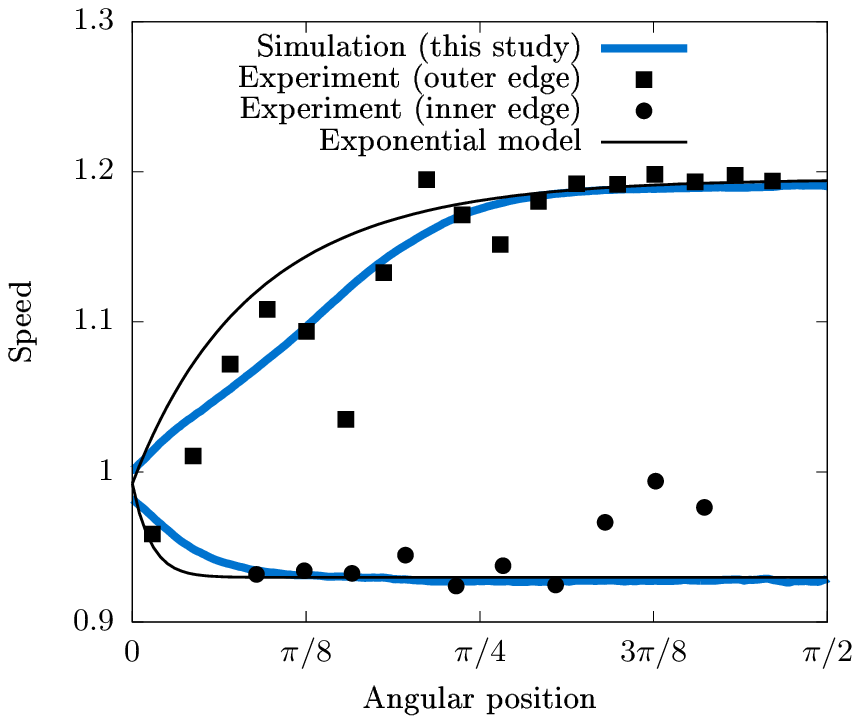}
    \caption{Detonation speeds at the inner and outer edge of the
      annular charge calculated in this numerical study along with
      experimental results and the exponential model suggested by
      Souers et al.~\cite{Souers1998}. The speeds are plotted against
      angular position $\theta$ defined in figure
      \ref{fig:lyle_configuration}.}
    \label{fig:lyle}
  \end{minipage}
\end{figure}

The detonation wave is initiated and propagates steadily in the straight section
at a constant speed. As it enters the annular arc every part of the front has
the same linear speed and hence the outer segments have lower angular velocity
compared to the inner parts. Thus, the inner parts propagate faster along the
arc. This results in the deformation of the shape of the front with its peak
shifted towards the inner wall. Figure \ref{fig:shock_fronts} shows the
evolution of the detonation front for the configuration of the Hayes
experiment. The curvature of the detonation front shape increases during the
transition period until steady state is reached, in which the front moves at
constant angular velocity and maintains its shape.

Figure \ref{fig:lyle} shows the numerical results for the speed of the
detonation wave along the inner and outer boundaries of the annular
charge against the experimental values and the exponential
time-dependent model suggested by Souers et al.~\cite{Souers1998}. The
inner part of the detonation moves to steady state quicker than the
outer part, which has a larger transition period, consistent with the
experimental results. The steady state values of the speeds are also
matched well. However, the numerical results do not follow the
exponential description of the transition phase. The speed at the
inner edge follows a slower exponential decay than predicted by the
model. At the outer edge, the detonation speed exhibits multiple
stages of distinct behavior. There is a short initial period where the
speed increase appears to be linear. Then the detonation exhibits
increasing acceleration and eventually, the acceleration decreases
until zero where the detonation has reached constant steady state
speed. This behavior is notably different to the exponential model and
is also within the scatter of the experimental points.

\subsection{The effect of the boundaries of the annular charge}
\label{subsect:inner_outer_effect}

The study of the Lyle configuration indicated that the evolution of
the detonation speed deviates from the suggested exponential model. To
investigate this outcome further and understand the processes involved
during the transition phase, we examine the effect of each of the two
boundaries of the annular charge separately.

\begin{figure}
  \centering
  \begin{minipage}{\linewidth}
    \includegraphics[scale=1]{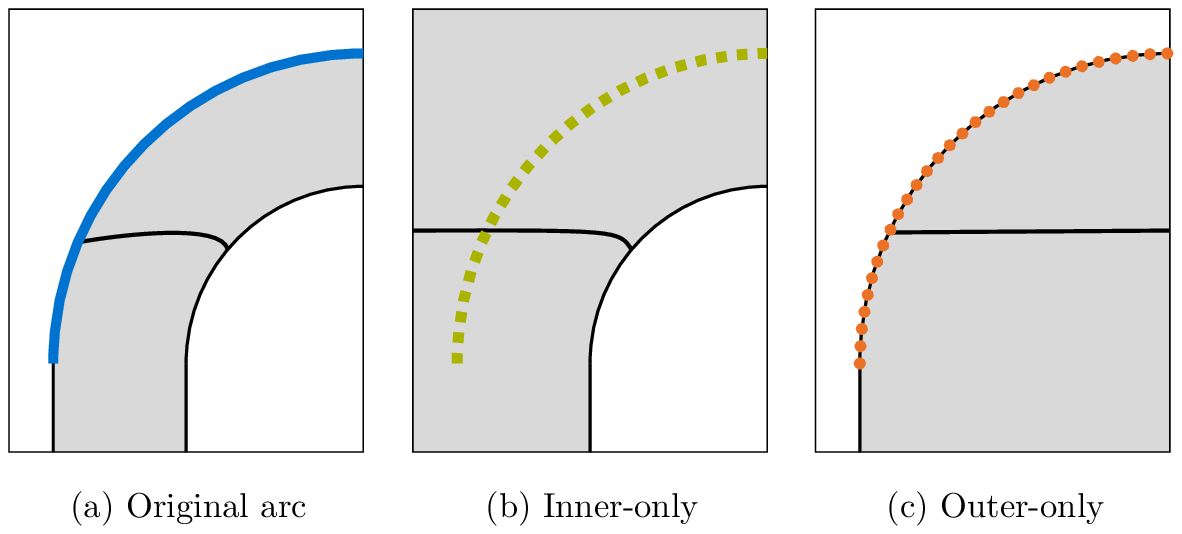}
    \caption{Depiction of the original annular configuration and of
      the test cases devised to examine the influence of the
      boundaries of the annulus. The shaded area represents the
      explosive and white space is air. Detonation speed measurements
      are made along the curves defined by the thick lines and are
      pattern and color coded to match the plots of figure
      \ref{fig:inner_outer_only_plot}.}
    \label{fig:inner_outer_only}
  \end{minipage}
  \begin{minipage}{\linewidth}
    \centering
    \includegraphics[scale=1]{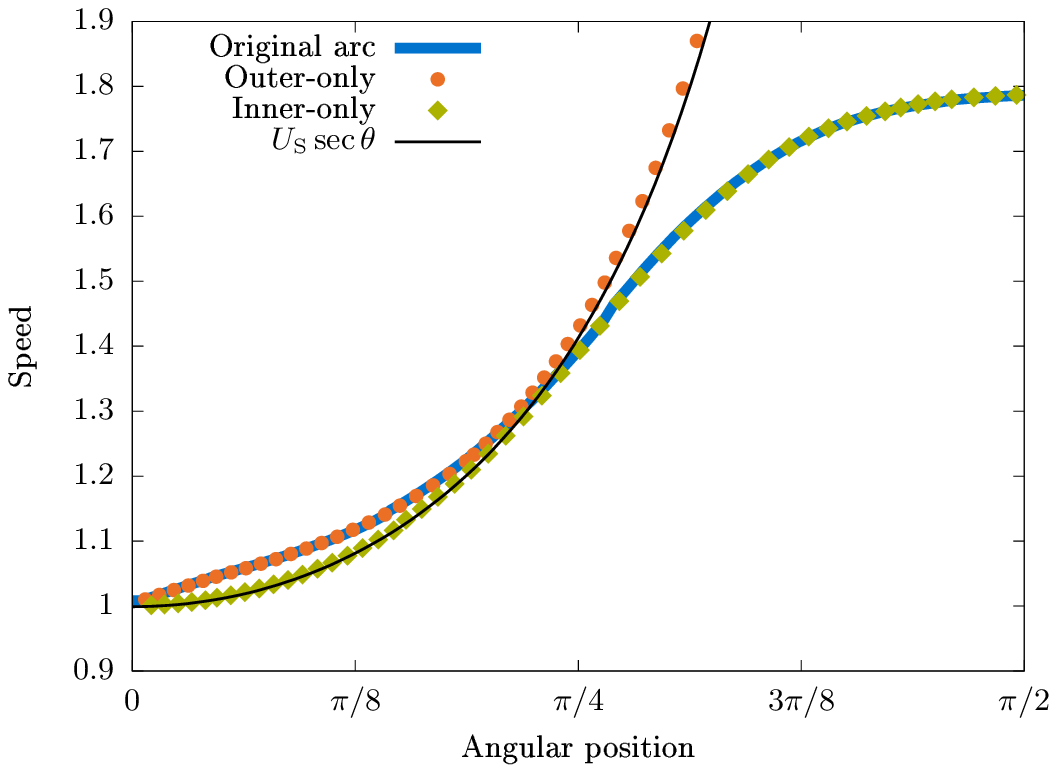}
    \caption{Detonation speed over angular position for the devised
      test cases and original annular arc setup. The configuration
      is of radius $R=6.616$ and width $D=6.616$.}
    \label{fig:inner_outer_only_plot}
  \end{minipage}
\end{figure}

We consider two test cases where one of the two boundaries of the
two-dimensional annular charge is removed and the remaining space is
filled with the explosive. This results in a semi-infinite explosive
charge with a single edge, as shown in the illustrations of figure
\ref{fig:inner_outer_only}. The configurations also include a straight
section where the detonation is initiated and left to reach steady
state. In both test cases, we calculate the speed of the detonation
wave along the curve defined by the outer boundary of the original arc
configuration. These curves are illustrated in figure
\ref{fig:inner_outer_only} and the resulting detonation speeds for
each test case are shown in figure \ref{fig:inner_outer_only_plot}.

In the case of inner-only boundary, the detonation along the outer
curve initially accelerates at an increasing rate. It then reaches an
inflection point and the acceleration decreases until it becomes zero
and the detonation travels at constant speed. The evolution of the
detonation speed in this case can be divided into two regimes based on
the physical processes involved. The change of geometry at the inner
edge alters the detonation dynamics and its effects travel along the
front at a finite speed. The first regime corresponds to the period
before the effects from the inner boundary reach the outer part of
the detonation front. An inflection point signals the start of the
second regime where the inner boundary effects have reached the outer
part and progressively move the detonation to steady state.

During the first regime, the outer part of the detonation is yet to
be affected by the changes in the geometry of the explosive charge and
the change in detonation speed is a result of local effects at the
curve representing the outer boundary of the original configuration. A
mathematical description of the first regime can be deduced by
considering that the detonation wave during this period is simply a
plane wave propagating in the direction normal to the front. Assuming
a constant straight section speed $U_{\op{S}}$, the speed measured
along the curve defined by the outer boundary $W$ will follow
\begin{equation}
  W = U_{\op{S}}\sec\theta,
  \label{eqn:sec_function}
\end{equation}
where $\theta$ is the angular position along the arc. Hence, the
measured speed is not the actual speed at which each part of the
detonation front propagates in the explosive, but rather the rate at
which the detonation front reaches the curve as it travels across
it.

In the second regime, the effects from the inner boundary reach the
outer part causing a change in the curvature of the detonation front
and in the direction of propagation. This results in the observed
decrease in the acceleration of the detonation which eventually
reaches steady state.

In the case of outer-only boundary, the speed exhibits an
initial stage where it appears to increase linearly. This is followed by
a stage of increasing acceleration of the detonation leading to very
high speeds. The detonation front is flat and travels at the same
direction as in the straight section.

We expect the propagation of the detonation wave in the outer-only
case to be similar to the first regime of the inner-only boundary
configuration because in both cases the front propagates at a constant
velocity and measurements are taken along a $90^\circ$ arc. Hence,
the detonation speed will follow equation
\eqref{eqn:sec_function}. This is indeed seen in the second stage, but
with an offset, because of the observed linear increase in detonation
speed during the initial stage.

The existence of the initial stage is attributed to the curvature of a
small segment of the detonation front next to the outer boundary of the
straight section. As discussed in section \ref{subsect:rate_sticks}
the front of a steady detonation in charges of finite size is curved
and its curvature depends on the diameter of the charge. In the case
of outer-only boundary, the explosive charge is semi-infinite and the
detonation front has a small curved segment only at the edge of the
charge while the rest is flat. Due to this convex curvature, the
front reaches the outer boundary faster than if it were flat,
resulting in the observed linear increase. The fact that this curvature
is limited to a small segment of the front next to the boundary means
that only a short initial stage is influenced by it.

The detonation speed along the outer boundary for the original annular
configuration is also shown in figure \ref{fig:inner_outer_only_plot}.  Based on
the test case of inner-only boundary, we again distinguish between two regimes
in which the change in detonation speed is caused by different effects. The
first regime is caused by local effects at the outer boundary of the annular
charge. This leads to the behavior seen in the case of outer-only boundary,
where the outer detonation speed depends on angular position; it increases
linearly at first and then as $\sec\theta$. The second regime is induced by the
effects of the inner edge. When these reach the outer edge of the annulus, the
outer speed goes through an inflection point and the acceleration of the
detonation decreases until it reaches constant steady state speed in a way that
exactly follows the second regime of the inner-only test case. This shows that
steady state is caused solely by the effects originating from the inner
edge. Similar findings have been reported by the asymptotic analysis of a DSD
model~\cite{Short2016}. In the thick arcs approximation $D/R \sim
\mathcal{O}(1)$, the dependence of steady angular velocity on inner radius and
degree of inner confinement is caused by a small boundary layer region along the
inner arc surface. Also, the outer radius and degree of confinement do not enter
in any terms that determine the steady angular velocity.

\subsection{The effect of the dimensions of the annular charge}

The dependence of detonation propagation on the dimensions of the
annular charge is investigated through a parametric study in which the
width and inner radius of the annulus are varied. A set of values
for the radius and width are selected as multiples of the greatest
common factor of the radius and width used in the Lyle and
Hayes experiments. The set of values used is shown in table \ref{tbl:param_study_values}.

\begin{table}
  \begin{center}
    \begin{tabular}{l@{\qquad}c@{\qquad}c@{\qquad}c@{\qquad}c@{\qquad}c@{\qquad}}
      \hline
      \hline
      Radius &  6.616 &  8.27 & 9.924 & 11.578 & 13.232 \\
      \hline
      Width & 1.654 & 3.308  & 4.962 & 6.616 \\
      \hline
      \hline
    \end{tabular}
  \end{center}
  \caption{The values for inner radius and width used in the parametric study to
    study the effect of the dimensions of the annulus}
  \label{tbl:param_study_values}
\end{table}

\begin{figure}
  \centering
  \includegraphics[width=\linewidth]{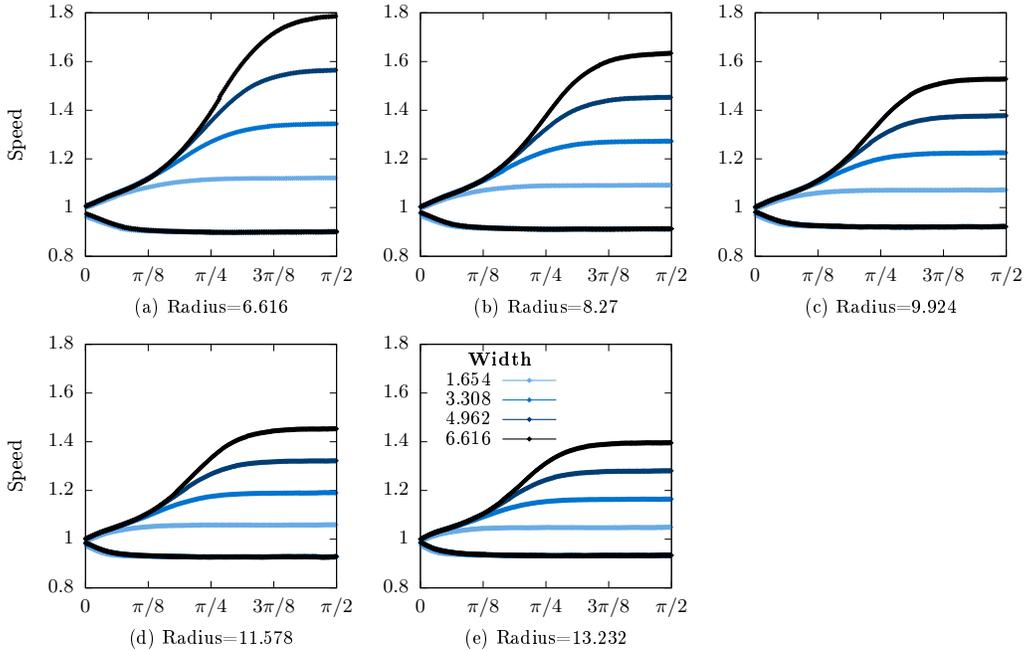}
  \caption{Detonation speeds over angular position along the inner and
    outer edge of the annular arc. The plots correspond to
    configurations of different radii in which the width varies from
    small (lighter color) to large (darker color).}
  \label{fig:param_radii}
\end{figure}

\begin{figure}
  \includegraphics[width=\linewidth]{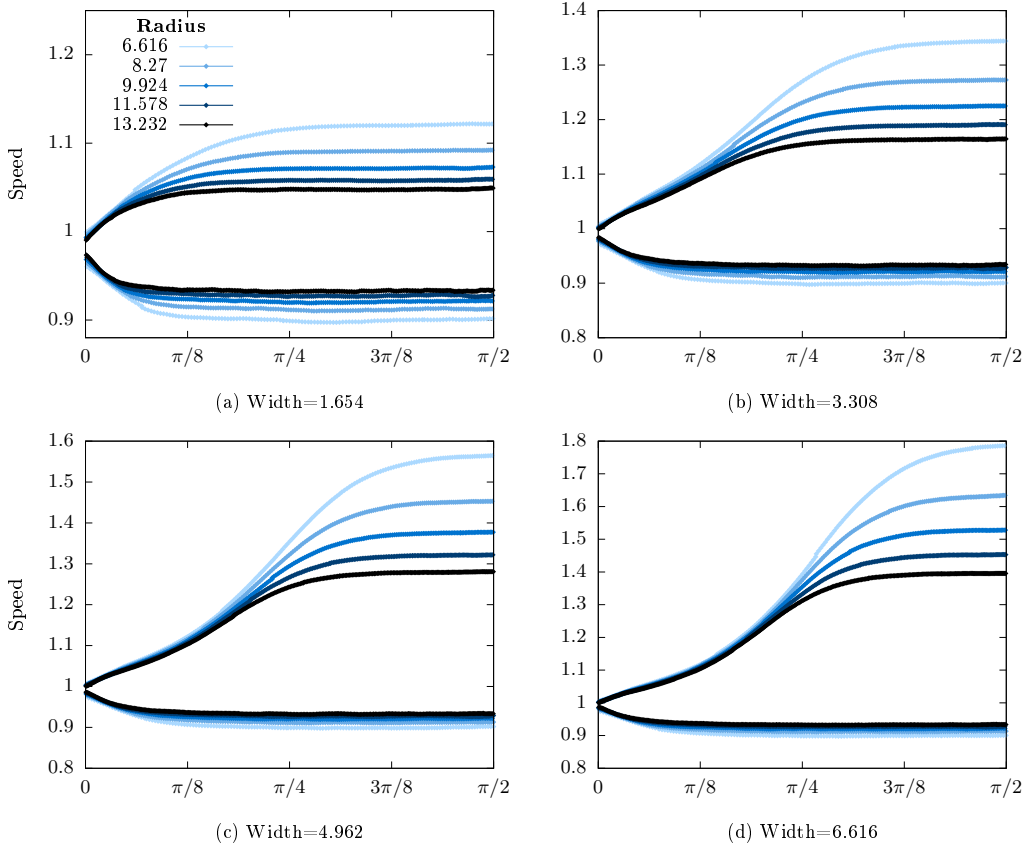}
  \caption{Detonation speeds over angular position along the inner and
    outer edge of the annular arc. The plots correspond to
    configurations of different width, in which the radius varies from
    small (lighter color) to large (darker color).}
  \label{fig:param_widths}
\end{figure}

Simulations were performed for all twenty combinations of the values
above. The obtained inner and outer speeds against angular position
along the annular section are presented in two sets of figures. Figure
\ref{fig:param_radii} arranges the results in configurations of the
same radius whereas the graphs of figure \ref{fig:param_widths}
correspond to configurations of the same width.

We note that stable detonation was observed in all configurations. Experiments
with gaseous explosives performed on similar configurations showed unsteady
propagation for certain configurations. However, the minimum width employed in
this study is double the failure radius of the considered explosive and any
unsteady or failing detonation was not observed.

\subsubsection{Steady state}
\label{subsect:param_steady_state}

We initially consider the steady state of detonation in annular
arcs and investigate its dependence on the dimensions of the explosive
charge. The plots of figure \ref{fig:param_radii} show a clear
increase of outer steady state speed $W_{\op{S}}$ with width. The
values are consistent with the condition of constant angular velocity
given by
\begin{equation}
  W_{\op{S}}  = \kappa(R,D) V_{\op{S}},
  \label{eqn:velocities_2}
\end{equation}
where $V_{\op{S}}$ is inner steady state speed and $\kappa$ is defined as the magnification coefficient which depends on the dimensions of the arc
\begin{equation}
  \kappa(R,D) = 1 + \frac{D}{R}.
  \label{eqn:magnification_coeff}
\end{equation}
In contrast, inner steady speeds do not differ for explosive
charges of different width and same radius. Despite the fact that
steady detonation speed in the straight section increases with width
due to the diameter effect, once the detonation reaches steady state
in the annular section, the inner speed is the same for all
configurations of the same radius. This implies that the steady state
angular velocity, $\omega_{\op{S}}$, is also independent of width since
it can be expressed as
\begin{equation}
  \omega_{\op{S}} \equiv \frac{V_{\op{S}}}{R}.
\end{equation}
The independence of steady angular velocity on the outer radius is also seen in
the asymptotic analysis of the DSD model performed by Short et
al.~\cite{Short2016} for arcs where $D/R \sim \mathcal{O}(1)$ as in this study.

\begin{figure}
  \centering
  \includegraphics[scale=1]{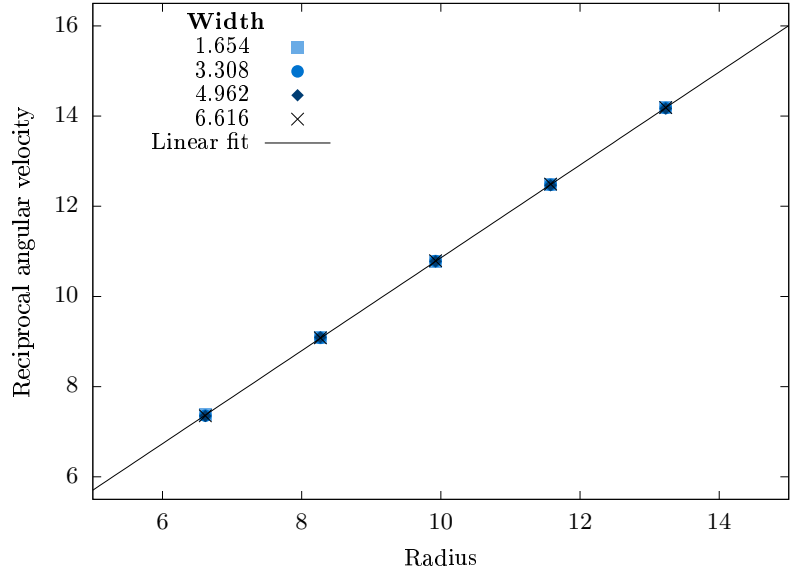}
  \caption{The reciprocal steady state angular velocity over inner
    radius for all studied configurations. These show an affine
    dependence of the reciprocal angular velocity on the inner radius
    of the annulus but no dependence on the width of the annulus.}
  \label{fig:steady_angular}
\end{figure}

The dependence of steady angular velocity on the dimensions of the
annulus is seen in figure \ref{fig:steady_angular}. Due to the
independence of the angular velocity on width, configurations of the
same width fall on the same point in the plot. Furthermore, we observe
an affine dependence of the reciprocal angular velocity on inner
radius which agrees with the results by Lubyatinsky et
al.~\cite{Lubyatinsky2004}. We perform a linear fit on the data using
the model function
\begin{equation}
  \omega_{\op{S}}^{-1}=\frac{R+\op{\Delta_0}}{\op{D_\infty}}, 
  \label{eqn:affine_fit}
\end{equation}
and obtain the values
\begin{align*}
  \op{D_\infty} &= \SI{0.9707 \pm 0.0011},\\
  \op{\Delta_0} &= \SI{0.537 \pm 0.012},
\end{align*}
which are characteristic of the combination of explosive and confining
material considered here. Parameter $\op{D_\infty}$ represents the
upper limit of detonation speed at the inner edge as
$R\rightarrow\infty$. Parameter $\op{\Delta_0}$ resembles the behavior
of failure thickness of a straight explosive slab because it decreases
with increasing impedance of the confining material as reported by
Lubyatinsky et al.~\cite{Lubyatinsky2004}.

Expression \eqref{eqn:affine_fit} allows for the
relations between inner or outer steady state speeds and the
dimensions of the annulus to be determined analytically. From figure
\ref{fig:param_widths} we can deduce that the inner steady state speed
increases with arc radius,
\begin{equation}
  \deriv{V_{\op{S}}}{R} > 0.
  \label{eqn:inner_over_r}
\end{equation}
Utilizing equation \eqref{eqn:affine_fit}, we see that the above
expression holds when $\op{\Delta_0}/\op{D_\infty}>0$, which has to be
satisfied to ensure positive angular velocity as $R\rightarrow0$.


The dependence of the outer steady state speed on the annulus dimensions
is also seen in the plots of figures \ref{fig:param_radii} and
\ref{fig:param_widths}. It increases with width
\begin{equation}
  \partderiv{W_{\op{S}}(R,D)}{D} > 0,
  \label{eqn:outer_over_d}
\end{equation}
and decreases with radius
\begin{equation}
  \partderiv{W_{\op{S}}(R,D)}{R} < 0,
  \label{eqn:outer_over_r}
\end{equation}
in contrast to the increase of steady inner speed
\eqref{eqn:inner_over_r}. Combining relation \eqref{eqn:affine_fit}
with the condition for constant angular velocity
\eqref{eqn:velocities_2} shows that expression
\eqref{eqn:outer_over_r} holds when width is larger than a threshold,
\begin{equation}
  D > \Delta_0.
  \label{eqn:condition}
\end{equation}
For the explosive and confining material considered here this value is
$\Delta_0=\SI{4.12}{\milli\metre}$ and the condition holds for all
configurations used in this study.

As expressed by \eqref{eqn:velocities_2}, outer speed depends on the
magnification coefficient and inner speed. These two quantities change
with radius in opposite ways. The magnification coefficient decreases
while the inner speed increases. Expression \eqref{eqn:condition} means
that when the width of the annulus is larger than a threshold, the
magnification part of equation \eqref{eqn:velocities_2} is dominant
and decreases more with radius than the inner speed increases, whereas
the opposite applies for widths smaller than the threshold.

\subsubsection{Transition phase}
\label{subsect:param_transition}

The transition phase is the period during which the detonation shifts
from a steady state of constant linear speed in the straight section
to a steady state of constant angular speed in the annular
section. The inner part of the detonation front reaches steady state
earlier than the outer edge of the front, which is the last segment to
reach steady state speed and determines the extent of the transition
phase.

The dimensions of the annulus influence the inner and outer speeds of
the detonation front during the transition phase in different
ways. Figure \ref{fig:param_radii} indicates that configurations of
different inner radius have qualitatively similar evolution of speeds
whereas figure \ref{fig:param_widths} shows distinct acceleration
profiles of the outer detonation front for configurations of different
width. In particular, configurations of larger widths demonstrate more
pronounced local effects at the outer boundary and more extensive
transition phases. The inner radius of the annulus only influences the
last stage of the transition phase and leads to reaching steady state
at smaller angles with increasing radius.

The extent of the transition phase is a function of the acceleration
profile during the transition phase and of the difference between the
steady speeds in the annular and straight section. Thus, its
dependence on the dimensions of the annulus can be deduced from
knowing the respective dependence of these two quantities. The dependence of steady
speeds on the dimensions of the annulus is well understood and is
presented in section \ref{subsect:param_steady_state}, but the
acceleration of the outer front is only known qualitatively, as
discussed in section \ref{subsect:inner_outer_effect} and no exact
function that describes the whole process is known.

We consider the extent of the transition phase in terms of angle and
time. We define the equilibration angle and time as the points at
which steady state is reached and thus they mark the transition phase
extent. Angle is measured as shown in figure
\ref{fig:lyle_configuration} and time is set to zero when the
detonation wave enters the annular region. The
criterion used to determine when steady state is reached is
\[ W \geq 0.99W_{\op{S}},\] where $W$ is the speed at the outer
edge. The value of steady outer speed $W_{\op{S}}$, is defined in
equation \eqref{eqn:velocities_2} and requires knowledge of the inner
steady state speed $V_{\op{S}}$. This value is obtained from the
numerical solution by averaging the speed values measured at the inner
edge well after it has reached steady state.

\begin{figure}
  \includegraphics[width=\linewidth]{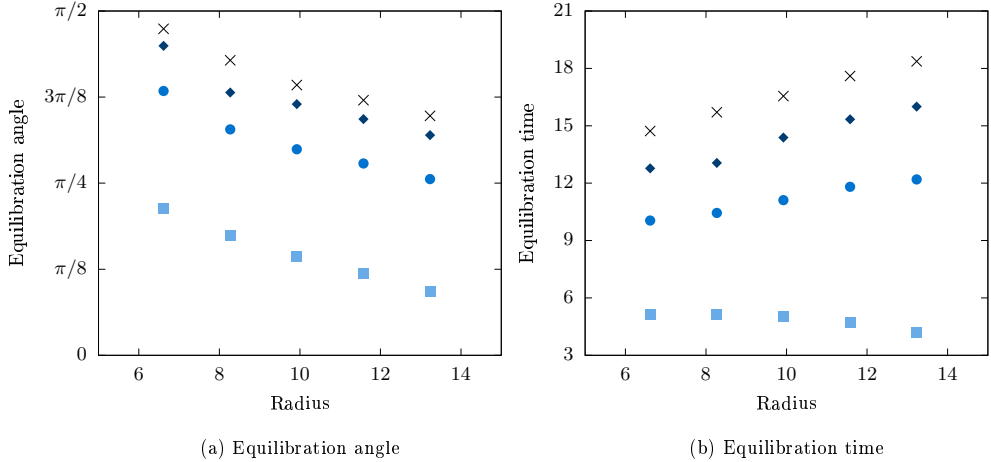}
  \caption{Equilibration angle and time for annular arcs of different
    inner radius and width. Time is zero when the detonation enters
    the annular arc. Configurations of different width are represented
    as seen in the legend of figure \ref{fig:steady_angular}.}
  \label{fig:equilibration}
\end{figure} 

Figure \ref{fig:equilibration} presents equilibration angles and times for the
studied configurations. The results indicate that both the equilibration angle
and time increase with the width of the explosive annulus. Charges of larger
width require more time for the effects of the inner edge to reach the outer
parts which results in a more extensive transition phase in terms of both angle
and time. In addition, the first regime is also more extensive and the
distinctive $\op{sec}\theta$ evolution is more prominent. In contrast,
configurations of small width have a less extensive first regime and transition
phase in general. In fact, for the configurations of the smallest width
considered (figure \ref{fig:param_widths}a), steady state is reached early
enough that the local effects of the outer edge do not develop significantly and
the transition to steady state can be described solely by a bounded growth
function.

The influence of the inner radius on the transition phase is more
complex. Equilibration angle decreases with radius for all
configurations but in the case of equilibration time we see opposing
behavior between configurations of different width. The exponential
time dependent model predicts that the transition phase duration
increases with steady outer speed. If this was a valid description of
the transition phase and using expression \eqref{eqn:outer_over_r},
the transition phase duration should decrease with inner
radius. Instead the equilibration time increases with radius, with the
exception of small widths.
 
\begin{figure}
  \includegraphics[width=\linewidth]{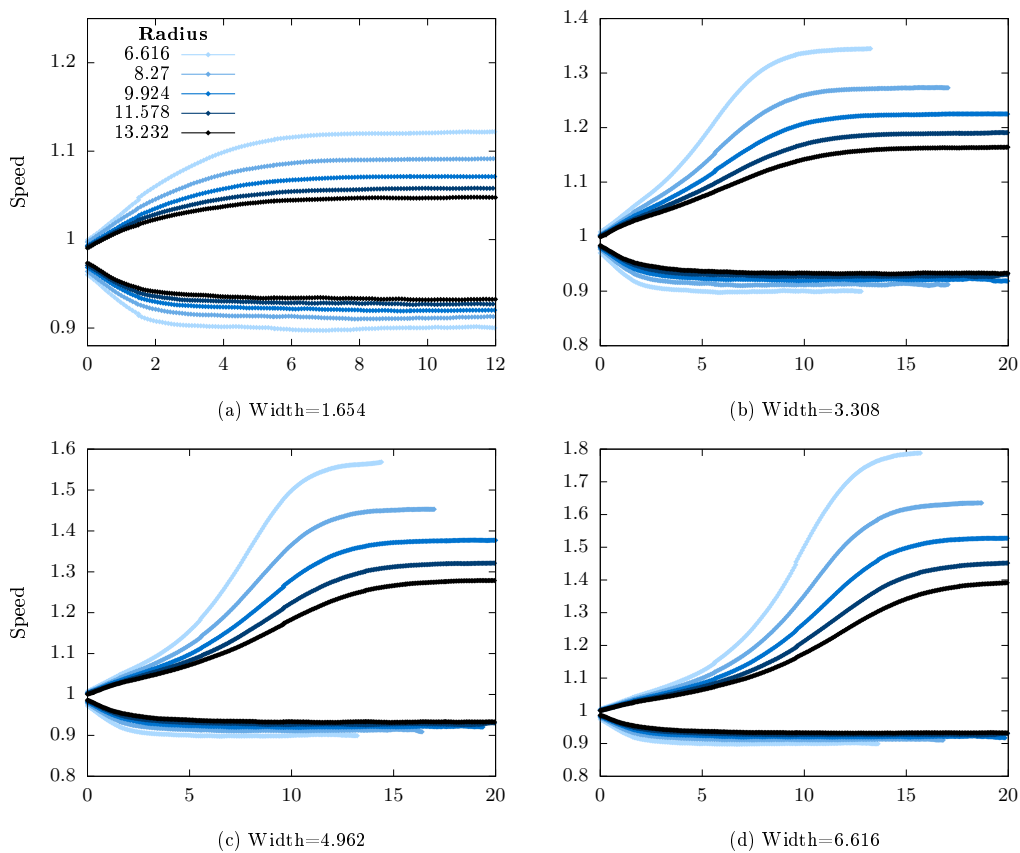}
  \caption{Detonation speeds over time along the inner and outer
    edge of the $90^\circ$ annular arc. The plots correspond to
    configurations of different width, in which the radius varies
    from small (lighter color) to large (darker color). Time is
    zero when the detonation wave enters the annular region.}
  \label{fig:param_widths_time}
\end{figure}

The increase of the transition duration despite the reduction of
steady speed indicates that the acceleration of the wave front
decreases with inner radius. As seen in section
\ref{subsect:inner_outer_effect}, the outer speed during the first
regime of the transition phase depends on angular position which
translates to a dependence on the dimensions of the annulus since,
\begin{equation}
  \deriv{W(\theta)}{t} = \deriv{W(\theta)}{\theta} \deriv{\theta}{t} = \deriv{W(\theta)}{\theta} \frac{W}{R+D}.
\end{equation}
The acceleration is inversely proportional to the
inner radius of the annulus and hence, large radii have slower
acceleration which results in reaching steady state at later times but
not at larger angles. This can also be seen in figure
\ref{fig:param_widths_time}, which shows detonation speeds over time
during the propagation of detonation in the annular section.

The discrepancy between equilibration angles and times is a result of
the local effects at the outer boundary. Thus, it is observed in
configurations of sufficiently large width, where the effects of the
outer edge are pronounced. If outer speed followed the suggested
time-dependent exponential function during the transition phase, the
transition phase duration would decrease with inner radius. This is
indeed seen in small width configurations in which the local effects
at the outer boundary are not developed and for which both equilibrium
angle and time decrease with radius.

\section{Conclusions}
\label{sect:conclusions}

This study is concerned with detonation propagation in condensed phase
explosive charges consisting of a straight and an annular
section. When a steady detonation in a straight charge enters the
annular section, it goes through a transition phase and eventually
reaches a new steady state of constant angular velocity. The
characteristic features of both phases of detonation propagation in
annular charges are identified and examined, as well as their
dependence on the dimensions of the annular arc.

A diffuse-interface formulation~\cite{Michael2016} is employed for the
calculations herein, which allows the modeling of a
two-phase explosive with distinct equations of state for the reactant
and products and of an additional inert material. The explosive
considered is LX-17 which is a granular, porous, polymer
bonded explosive. It is modeled by two JWL equations of state and the
Ignition and Growth reaction rate law. This provides a macroscopic
description of the effects of the micro-structure of a heterogeneous
high explosive and enables the homogeneous treatment of the
explosive.

The computations in this study were performed within a parallel
adaptive mesh refinement framework which allows the use of high
resolution with small computational cost. Convergence studies were
performed for all configurations studied to ensure that the solution
is independent of the grid resolution.

The mathematical model and numerical methods were validated through
several tests. These included the study of steady detonation in
one-dimensional and cylindrical charges. The structure of the steady
detonation obtained from the numerical solution was consistent with
theory and was compared against analytic solutions in the
one-dimensional case. In addition, a study of detonation speed over
radius in cylindrical charges was performed. This showed good agreement
with experimental values and demonstrated the capability of the model to
capture the diameter effect curve of the explosive.

The study of annular charges follows the configurations used in the
Lyle and Hayes experiments~\cite{Souers1998}. The steady state
detonation speeds show good agreement with the experimental values but
the transition phase deviates from the suggested exponential
model. The numerical solution of the transition phase indicates that
the outer speed increases linearly at the beginning, then at an
increasing rate and in the final stage the acceleration decreases to
zero and the detonation reaches steady state.

The effects that govern the evolution of detonation speed during the
transition phase are investigated through configurations with
only one of the boundaries that make up the annular arc. These
indicate that steady state is induced by effects originating from the
inner edge and travel along the front at a finite speed. Thus, the transition
phase of the outer speed can be divided into two regimes. In the first
regime, the effects of the inner edge have not yet reached the outer
part of the detonation and the outer speed is governed by local
effects at the outer edge. These lead to a dependence of the
detonation speed on the angular position along the arc. The second
regime begins when the effects from the inner edge reach the outer part
of the detonation. These change the curvature of the detonation front
and lead the detonation speed towards the steady state value in a bounded
growth manner.
 
The dependence of detonation propagation on the dimensions of the
annular charge was investigated through a parametric study. We varied the
inner radius and width of the charge and obtained the corresponding
detonation speeds along the annulus. Results show that steady angular
velocity depends only on the inner radius. In particular we observed
an affine dependence of steady state angular velocity on the inner
radius which has also been reported by Lubyatinsky et
al.~\cite{Lubyatinsky2004}. The width of the annulus does not
influence the steady angular velocity but it affects the outer steady
speed which increases with width due to the condition of constant
angular velocity.

The dimensions of the annulus influence the transition phase as
well. The width of the charge determines the extent of the first
regime of the transition phase with larger widths leading to more
pronounced local effects at the outer boundary. The inner radius
influences only the second regime of the transition phase. Increasing
radius brings the shape of the charge closer to a straight charge and
results in less difference between inner and outer steady detonation
speeds, as well as reduced angles at which steady state is reached.

The extent of the transition phase was studied in terms of the angle
and time at which steady state is reached. Both equilibration time
and angle increase with width for configurations of the same inner
radius. For annuli of the same width, increasing inner radius leads to
a decrease in equilibration angle. However, the equilibration time shows
different behavior depending on the width of the configuration. It
increases with radius for large widths and decreases for the
configuration of the smallest width.

This discrepancy is attributed to the first regime of the transition
phase. The dependence of detonation outer speed on angular position
during this regime leads to an inversely proportional relation between the
acceleration of the outer part of the detonation and the inner radius
of the annulus. This results in longer transition duration as
the inner radius is increased in contrast to the decrease in equilibration
angle. In the configurations of the smallest width, the equilibration
time decreases with inner radius because the first regime is small and
does not develop sufficiently to influence the transition duration.

%
%

%

\begin{acknowledgments}
  This work was supported by the UK Engineering and Physical Sciences
  Research Council (EPSRC) grant 1498435 for the University of
  Cambridge and by Orica Mining Services.
\end{acknowledgments}

\bibliography{references.bib}

\begin{thebibliography}{27}%
\makeatletter
\providecommand \@ifxundefined [1]{%
 \@ifx{#1\undefined}
}%
\providecommand \@ifnum [1]{%
 \ifnum #1\expandafter \@firstoftwo
 \else \expandafter \@secondoftwo
 \fi
}%
\providecommand \@ifx [1]{%
 \ifx #1\expandafter \@firstoftwo
 \else \expandafter \@secondoftwo
 \fi
}%
\providecommand \natexlab [1]{#1}%
\providecommand \enquote  [1]{``#1''}%
\providecommand \bibnamefont  [1]{#1}%
\providecommand \bibfnamefont [1]{#1}%
\providecommand \citenamefont [1]{#1}%
\providecommand \href@noop [0]{\@secondoftwo}%
\providecommand \href [0]{\begingroup \@sanitize@url \@href}%
\providecommand \@href[1]{\@@startlink{#1}\@@href}%
\providecommand \@@href[1]{\endgroup#1\@@endlink}%
\providecommand \@sanitize@url [0]{\catcode `\\12\catcode `\$12\catcode
  `\&12\catcode `\#12\catcode `\^12\catcode `\_12\catcode `\%12\relax}%
\providecommand \@@startlink[1]{}%
\providecommand \@@endlink[0]{}%
\providecommand \url  [0]{\begingroup\@sanitize@url \@url }%
\providecommand \@url [1]{\endgroup\@href {#1}{\urlprefix }}%
\providecommand \urlprefix  [0]{URL }%
\providecommand \Eprint [0]{\href }%
\providecommand \doibase [0]{http://dx.doi.org/}%
\providecommand \selectlanguage [0]{\@gobble}%
\providecommand \bibinfo  [0]{\@secondoftwo}%
\providecommand \bibfield  [0]{\@secondoftwo}%
\providecommand \translation [1]{[#1]}%
\providecommand \BibitemOpen [0]{}%
\providecommand \bibitemStop [0]{}%
\providecommand \bibitemNoStop [0]{.\EOS\space}%
\providecommand \EOS [0]{\spacefactor3000\relax}%
\providecommand \BibitemShut  [1]{\csname bibitem#1\endcsname}%
\let\auto@bib@innerbib\@empty
\bibitem [{\citenamefont {Souers}\ \emph {et~al.}(1998)\citenamefont {Souers},
  \citenamefont {Anderson}, \citenamefont {Hayes}, \citenamefont {Lyle},
  \citenamefont {Lee}, \citenamefont {McGuire},\ and\ \citenamefont
  {Tarver}}]{Souers1998}%
  \BibitemOpen
  \bibfield  {author} {\bibinfo {author} {\bibfnamefont {P.~C.}\ \bibnamefont
  {Souers}}, \bibinfo {author} {\bibfnamefont {S.~R.}\ \bibnamefont
  {Anderson}}, \bibinfo {author} {\bibfnamefont {B.}~\bibnamefont {Hayes}},
  \bibinfo {author} {\bibfnamefont {J.}~\bibnamefont {Lyle}}, \bibinfo {author}
  {\bibfnamefont {E.~L.}\ \bibnamefont {Lee}}, \bibinfo {author} {\bibfnamefont
  {S.~M.}\ \bibnamefont {McGuire}}, \ and\ \bibinfo {author} {\bibfnamefont
  {C.~M.}\ \bibnamefont {Tarver}},\ }\bibfield  {title} {\enquote {\bibinfo
  {title} {{Corner turning rib tests on LX-17}},}\ }\href {\doibase
  10.1002/(SICI)1521-4087(199808)23:4<200::AID-PREP200>3.0.CO;2-9} {\bibfield
  {journal} {\bibinfo  {journal} {Propellants Explosives Pyrotechnics}\
  }\textbf {\bibinfo {volume} {23}},\ \bibinfo {pages} {200--207} (\bibinfo
  {year} {1998})}\BibitemShut {NoStop}%
\bibitem [{\citenamefont {{Lubyatinsky, S. N. Batalov, S. V. Garmashev, A. Yu.
  Israelyan, V. G. Kostitsyn, O. V. Loboiko, B. G. Pashentsev, V. A. Sibilev,
  V. A. Smirnov, E. B. Filin}}(2004)}]{Lubyatinsky2004}%
  \BibitemOpen
  \bibfield  {author} {\bibinfo {author} {\bibfnamefont {V.~P.}\ \bibnamefont
  {{Lubyatinsky, S. N. Batalov, S. V. Garmashev, A. Yu. Israelyan, V. G.
  Kostitsyn, O. V. Loboiko, B. G. Pashentsev, V. A. Sibilev, V. A. Smirnov, E.
  B. Filin}}},\ }\bibfield  {title} {\enquote {\bibinfo {title} {{Detonation
  Propagation in 180° Ribs of an Insensitive High Explosive}},}\ }in\ \href
  {\doibase http://dx.doi.org/10.1063/1.17803721.1780372} {\emph {\bibinfo
  {booktitle} {AIP Conference Proceedings}}},\ Vol.\ \bibinfo {volume} {706},\
  \bibinfo {organization} {AIP Publishing}\ (\bibinfo  {publisher} {AIP},\
  \bibinfo {year} {2004})\ pp.\ \bibinfo {pages} {859--862}\BibitemShut
  {NoStop}%
\bibitem [{\citenamefont {Tonghu}\ \emph {et~al.}(1998)\citenamefont {Tonghu},
  \citenamefont {Qingzhong}, \citenamefont {Feng}, \citenamefont {Lishi},
  \citenamefont {Zhi},\ and\ \citenamefont {Wen}}]{Tonghu1998}%
  \BibitemOpen
  \bibfield  {author} {\bibinfo {author} {\bibfnamefont {Z.}~\bibnamefont
  {Tonghu}}, \bibinfo {author} {\bibfnamefont {L.}~\bibnamefont {Qingzhong}},
  \bibinfo {author} {\bibfnamefont {Z.}~\bibnamefont {Feng}}, \bibinfo {author}
  {\bibfnamefont {H.}~\bibnamefont {Lishi}}, \bibinfo {author} {\bibfnamefont
  {H.}~\bibnamefont {Zhi}}, \ and\ \bibinfo {author} {\bibfnamefont
  {G.}~\bibnamefont {Wen}},\ }\bibfield  {title} {\enquote {\bibinfo {title}
  {{An Experimental Study of Detonation Propagation in the Arc Insensitive High
  Explosive Initiated on the Basal Plane}},}\ }in\ \href@noop {} {\emph
  {\bibinfo {booktitle} {Eleventh International Detonation Symposium, Snowmass,
  Colorado, USA}}}\ (\bibinfo {year} {1998})\ pp.\ \bibinfo {pages}
  {1023--1028}\BibitemShut {NoStop}%
\bibitem [{\citenamefont {Bdzil}\ \emph {et~al.}(2003)\citenamefont {Bdzil},
  \citenamefont {Aslam}, \citenamefont {Henninger},\ and\ \citenamefont
  {Quirk}}]{Bdzil2003}%
  \BibitemOpen
  \bibfield  {author} {\bibinfo {author} {\bibfnamefont {J.~B.}\ \bibnamefont
  {Bdzil}}, \bibinfo {author} {\bibfnamefont {T.}~\bibnamefont {Aslam}},
  \bibinfo {author} {\bibfnamefont {T.~D.}\ \bibnamefont {Henninger}}, \ and\
  \bibinfo {author} {\bibfnamefont {J.~J.}\ \bibnamefont {Quirk}},\ }\bibfield
  {title} {\enquote {\bibinfo {title} {{High-Explosives Performance}},}\
  }\href@noop {} {\bibfield  {journal} {\bibinfo  {journal} {Los Alamos
  Science}\ }\textbf {\bibinfo {volume} {28}},\ \bibinfo {pages} {96--110}
  (\bibinfo {year} {2003})}\BibitemShut {NoStop}%
\bibitem [{\citenamefont {Hill}\ and\ \citenamefont {Aslam}(2010)}]{Hill2010}%
  \BibitemOpen
  \bibfield  {author} {\bibinfo {author} {\bibfnamefont {L.~G.}\ \bibnamefont
  {Hill}}\ and\ \bibinfo {author} {\bibfnamefont {T.~D.}\ \bibnamefont
  {Aslam}},\ }\bibfield  {title} {\enquote {\bibinfo {title} {{Detonation shock
  dynamics calibration for PBX 9502 with temperature, density, and material lot
  variations}},}\ }in\ \href@noop {} {\emph {\bibinfo {booktitle} {Fourteenth
  International Detonation Symposium}}}\ (\bibinfo {year} {2010})\ pp.\
  \bibinfo {pages} {779----788}\BibitemShut {NoStop}%
\bibitem [{\citenamefont {Short}\ \emph {et~al.}(2016)\citenamefont {Short},
  \citenamefont {Quirk}, \citenamefont {Meyer},\ and\ \citenamefont
  {Chiquete}}]{Short2016}%
  \BibitemOpen
  \bibfield  {author} {\bibinfo {author} {\bibfnamefont {M.}~\bibnamefont
  {Short}}, \bibinfo {author} {\bibfnamefont {J.~J.}\ \bibnamefont {Quirk}},
  \bibinfo {author} {\bibfnamefont {C.~D.}\ \bibnamefont {Meyer}}, \ and\
  \bibinfo {author} {\bibfnamefont {C.}~\bibnamefont {Chiquete}},\ }\bibfield
  {title} {\enquote {\bibinfo {title} {Steady detonation propagation in a
  circular arc: a detonation shock dynamics model},}\ }\href {\doibase
  10.1017/jfm.2016.597} {\bibfield  {journal} {\bibinfo  {journal} {Journal of
  Fluid Mechanics}\ }\textbf {\bibinfo {volume} {807}},\ \bibinfo {pages}
  {87--134} (\bibinfo {year} {2016})}\BibitemShut {NoStop}%
\bibitem [{\citenamefont {Nakayama}\ \emph {et~al.}(2012)\citenamefont
  {Nakayama}, \citenamefont {Moriya}, \citenamefont {Kasahara}, \citenamefont
  {Matsuo}, \citenamefont {Sasamoto},\ and\ \citenamefont
  {Funaki}}]{Nakayama2012}%
  \BibitemOpen
  \bibfield  {author} {\bibinfo {author} {\bibfnamefont {H.}~\bibnamefont
  {Nakayama}}, \bibinfo {author} {\bibfnamefont {T.}~\bibnamefont {Moriya}},
  \bibinfo {author} {\bibfnamefont {J.}~\bibnamefont {Kasahara}}, \bibinfo
  {author} {\bibfnamefont {A.}~\bibnamefont {Matsuo}}, \bibinfo {author}
  {\bibfnamefont {Y.}~\bibnamefont {Sasamoto}}, \ and\ \bibinfo {author}
  {\bibfnamefont {I.}~\bibnamefont {Funaki}},\ }\bibfield  {title} {\enquote
  {\bibinfo {title} {Stable detonation wave propagation in
  rectangular-cross-section curved channels},}\ }\href {\doibase
  10.1016/j.combustflame.2011.07.022} {\bibfield  {journal} {\bibinfo
  {journal} {Combustion and flame}\ }\textbf {\bibinfo {volume} {159}},\
  \bibinfo {pages} {859--869} (\bibinfo {year} {2012})}\BibitemShut {NoStop}%
\bibitem [{\citenamefont {Souers}\ \emph {et~al.}(2002)\citenamefont {Souers},
  \citenamefont {McGuire}, \citenamefont {Garza}, \citenamefont {Roeske},\ and\
  \citenamefont {Vitello}}]{Souers2002}%
  \BibitemOpen
  \bibfield  {author} {\bibinfo {author} {\bibfnamefont {P.~C.}\ \bibnamefont
  {Souers}}, \bibinfo {author} {\bibfnamefont {E.}~\bibnamefont {McGuire}},
  \bibinfo {author} {\bibfnamefont {R.~G.}\ \bibnamefont {Garza}}, \bibinfo
  {author} {\bibfnamefont {F.}~\bibnamefont {Roeske}}, \ and\ \bibinfo {author}
  {\bibfnamefont {P.}~\bibnamefont {Vitello}},\ }\bibfield  {title} {\enquote
  {\bibinfo {title} {{The Diverging Sphere and the Rib in Prompt
  Detonation}},}\ }\href@noop {} {\bibfield  {journal} {\bibinfo  {journal}
  {12th Symposium (International) on Detonation}\ }\textbf {\bibinfo {volume}
  {5}},\ \bibinfo {pages} {1--7} (\bibinfo {year} {2002})}\BibitemShut
  {NoStop}%
\bibitem [{\citenamefont {V{\'{a}}genknecht}\ and\ \citenamefont
  {Adam{\'{i}}k}(2006)}]{Vagenknecht2006}%
  \BibitemOpen
  \bibfield  {author} {\bibinfo {author} {\bibfnamefont {J.}~\bibnamefont
  {V{\'{a}}genknecht}}\ and\ \bibinfo {author} {\bibfnamefont {V.}~\bibnamefont
  {Adam{\'{i}}k}},\ }\bibfield  {title} {\enquote {\bibinfo {title} {{A
  contribution to the analysis of corner turning rib problems}},}\ }\href
  {\doibase 10.1002/prep.200600041} {\bibfield  {journal} {\bibinfo  {journal}
  {Propellants, Explosives, Pyrotechnics}\ }\textbf {\bibinfo {volume} {31}},\
  \bibinfo {pages} {299--305} (\bibinfo {year} {2006})}\BibitemShut {NoStop}%
\bibitem [{\citenamefont {Tarver}\ \emph {et~al.}(2008)\citenamefont {Tarver},
  \citenamefont {Chidester}, \citenamefont {Elert}, \citenamefont {Furnish},
  \citenamefont {Chau}, \citenamefont {Holmes},\ and\ \citenamefont
  {Nguyen}}]{Tarver2007}%
  \BibitemOpen
  \bibfield  {author} {\bibinfo {author} {\bibfnamefont {C.~M.}\ \bibnamefont
  {Tarver}}, \bibinfo {author} {\bibfnamefont {S.~K.}\ \bibnamefont
  {Chidester}}, \bibinfo {author} {\bibfnamefont {M.}~\bibnamefont {Elert}},
  \bibinfo {author} {\bibfnamefont {M.~D.}\ \bibnamefont {Furnish}}, \bibinfo
  {author} {\bibfnamefont {R.}~\bibnamefont {Chau}}, \bibinfo {author}
  {\bibfnamefont {N.}~\bibnamefont {Holmes}}, \ and\ \bibinfo {author}
  {\bibfnamefont {J.}~\bibnamefont {Nguyen}},\ }\bibfield  {title} {\enquote
  {\bibinfo {title} {{Ignition and growth modeling of detonating TATB cones and
  arcs}},}\ }in\ \href {\doibase 10.1063/1.2833085} {\emph {\bibinfo
  {booktitle} {AIP Conference Proceedings}}},\ Vol.\ \bibinfo {volume} {955}\
  (\bibinfo {year} {2008})\ pp.\ \bibinfo {pages} {429--432}\BibitemShut
  {NoStop}%
\bibitem [{\citenamefont {Michael}\ and\ \citenamefont
  {Nikiforakis}(2016)}]{Michael2016}%
  \BibitemOpen
  \bibfield  {author} {\bibinfo {author} {\bibfnamefont {L.}~\bibnamefont
  {Michael}}\ and\ \bibinfo {author} {\bibfnamefont {N.}~\bibnamefont
  {Nikiforakis}},\ }\bibfield  {title} {\enquote {\bibinfo {title} {{A hybrid
  formulation for the numerical simulation of condensed phase explosives}},}\
  }\href {\doibase 10.1016/j.jcp.2016.04.017} {\bibfield  {journal} {\bibinfo
  {journal} {Journal of Computational Physics}\ }\textbf {\bibinfo {volume}
  {316}},\ \bibinfo {pages} {193--217} (\bibinfo {year} {2016})}\BibitemShut
  {NoStop}%
\bibitem [{\citenamefont {Lee}\ and\ \citenamefont {Tarver}(1980)}]{Lee1980}%
  \BibitemOpen
  \bibfield  {author} {\bibinfo {author} {\bibfnamefont {E.~L.}\ \bibnamefont
  {Lee}}\ and\ \bibinfo {author} {\bibfnamefont {C.~M.}\ \bibnamefont
  {Tarver}},\ }\bibfield  {title} {\enquote {\bibinfo {title}
  {{Phenomenological model of shock initiation in heterogeneous explosives}},}\
  }\href {\doibase 10.1063/1.862940} {\bibfield  {journal} {\bibinfo  {journal}
  {Physics of Fluids}\ }\textbf {\bibinfo {volume} {23}},\ \bibinfo {pages}
  {2362} (\bibinfo {year} {1980})}\BibitemShut {NoStop}%
\bibitem [{\citenamefont {Allaire}, \citenamefont {Clerc},\ and\ \citenamefont
  {Kokh}(2002)}]{Allaire2002}%
  \BibitemOpen
  \bibfield  {author} {\bibinfo {author} {\bibfnamefont {G.}~\bibnamefont
  {Allaire}}, \bibinfo {author} {\bibfnamefont {S.}~\bibnamefont {Clerc}}, \
  and\ \bibinfo {author} {\bibfnamefont {S.}~\bibnamefont {Kokh}},\ }\bibfield
  {title} {\enquote {\bibinfo {title} {{A Five-Equation Model for the
  Simulation of Interfaces between Compressible Fluids}},}\ }\href {\doibase
  10.1006/jcph.2002.7143} {\bibfield  {journal} {\bibinfo  {journal} {Journal
  of Computational Physics}\ }\textbf {\bibinfo {volume} {181}},\ \bibinfo
  {pages} {577--616} (\bibinfo {year} {2002})}\BibitemShut {NoStop}%
\bibitem [{\citenamefont {Banks}\ \emph {et~al.}(2007)\citenamefont {Banks},
  \citenamefont {Schwendeman}, \citenamefont {Kapila},\ and\ \citenamefont
  {Henshaw}}]{Banks2007}%
  \BibitemOpen
  \bibfield  {author} {\bibinfo {author} {\bibfnamefont {J.~W.}\ \bibnamefont
  {Banks}}, \bibinfo {author} {\bibfnamefont {D.~W.}\ \bibnamefont
  {Schwendeman}}, \bibinfo {author} {\bibfnamefont {A.~K.}\ \bibnamefont
  {Kapila}}, \ and\ \bibinfo {author} {\bibfnamefont {W.~D.}\ \bibnamefont
  {Henshaw}},\ }\bibfield  {title} {\enquote {\bibinfo {title} {{A
  high-resolution Godunov method for compressible multi-material flow on
  overlapping grids}},}\ }\href {\doibase 10.1016/j.jcp.2006.09.014} {\bibfield
   {journal} {\bibinfo  {journal} {Journal of Computational Physics}\ }\textbf
  {\bibinfo {volume} {223}},\ \bibinfo {pages} {262--297} (\bibinfo {year}
  {2007})}\BibitemShut {NoStop}%
\bibitem [{\citenamefont {Kapila}\ \emph {et~al.}(2007)\citenamefont {Kapila},
  \citenamefont {Schwendeman}, \citenamefont {Bdzil},\ and\ \citenamefont
  {Henshaw}}]{Kapila2007}%
  \BibitemOpen
  \bibfield  {author} {\bibinfo {author} {\bibfnamefont {A.~K.}\ \bibnamefont
  {Kapila}}, \bibinfo {author} {\bibfnamefont {D.~W.}\ \bibnamefont
  {Schwendeman}}, \bibinfo {author} {\bibfnamefont {J.~B.}\ \bibnamefont
  {Bdzil}}, \ and\ \bibinfo {author} {\bibfnamefont {W.~D.}\ \bibnamefont
  {Henshaw}},\ }\bibfield  {title} {\enquote {\bibinfo {title} {{A study of
  detonation diffraction in the ignition-and-growth model}},}\ }\href {\doibase
  10.1080/13647830701235774} {\bibfield  {journal} {\bibinfo  {journal}
  {Combustion Theory and Modelling}\ }\textbf {\bibinfo {volume} {11}},\
  \bibinfo {pages} {781--822} (\bibinfo {year} {2007})}\BibitemShut {NoStop}%
\bibitem [{\citenamefont {Tarver}(2005)}]{Tarver2005}%
  \BibitemOpen
  \bibfield  {author} {\bibinfo {author} {\bibfnamefont {C.~M.}\ \bibnamefont
  {Tarver}},\ }\bibfield  {title} {\enquote {\bibinfo {title} {{Ignition and
  Growth Modeling of LX-17 Hockey Puck Experiments}},}\ }\href {\doibase
  10.1002/prep.200400092} {\bibfield  {journal} {\bibinfo  {journal}
  {Propellants, Explosives, Pyrotechnics}\ }\textbf {\bibinfo {volume} {30}},\
  \bibinfo {pages} {109--117} (\bibinfo {year} {2005})}\BibitemShut {NoStop}%
\bibitem [{\citenamefont {Tarver}(2010)}]{Tarver2010}%
  \BibitemOpen
  \bibfield  {author} {\bibinfo {author} {\bibfnamefont {C.~M.}\ \bibnamefont
  {Tarver}},\ }\bibfield  {title} {\enquote {\bibinfo {title} {{Corner Turning
  and Shock Desensitization Experiments plus Numerical Modeling of Detonation
  Waves in the Triaminotrinitrobenzene Based Explosive LX-17}},}\ }\href
  {\doibase 10.1021/jp9098733} {\bibfield  {journal} {\bibinfo  {journal} {The
  Journal of Physical Chemistry A}\ }\textbf {\bibinfo {volume} {114}},\
  \bibinfo {pages} {2727--2736} (\bibinfo {year} {2010})}\BibitemShut {NoStop}%
\bibitem [{\citenamefont {Fickett}\ and\ \citenamefont
  {Davis}(1973)}]{Fickett2000}%
  \BibitemOpen
  \bibfield  {author} {\bibinfo {author} {\bibfnamefont {W.}~\bibnamefont
  {Fickett}}\ and\ \bibinfo {author} {\bibfnamefont {W.~C.}\ \bibnamefont
  {Davis}},\ }\href@noop {} {\emph {\bibinfo {title} {{Detonation: Theory and
  Experiment}}}}\ (\bibinfo  {publisher} {Dover Publications},\ \bibinfo {year}
  {1973})\ p.\ \bibinfo {pages} {363}\BibitemShut {NoStop}%
\bibitem [{\citenamefont {Lee}(2008)}]{Lee2008}%
  \BibitemOpen
  \bibfield  {author} {\bibinfo {author} {\bibfnamefont {J.~H.}\ \bibnamefont
  {Lee}},\ }\href {\doibase 10.2514/1.43659} {\emph {\bibinfo {title} {{The
  detonation phenomenon}}}},\ Vol.~\bibinfo {volume} {2}\ (\bibinfo
  {publisher} {Cambridge University Press Cambridge},\ \bibinfo {year} {2008})\
  p.\ \bibinfo {pages} {402}\BibitemShut {NoStop}%
\bibitem [{\citenamefont {Van~Leer}(1997)}]{vanLeer1997}%
  \BibitemOpen
  \bibfield  {author} {\bibinfo {author} {\bibfnamefont {B.}~\bibnamefont
  {Van~Leer}},\ }\bibfield  {title} {\enquote {\bibinfo {title} {On the
  relation between the upwind-differencing schemes of godunov, engquist—osher
  and roe},}\ }in\ \href@noop {} {\emph {\bibinfo {booktitle} {Upwind and
  high-Resolution schemes}}}\ (\bibinfo  {publisher} {Springer},\ \bibinfo
  {year} {1997})\ pp.\ \bibinfo {pages} {33--52}\BibitemShut {NoStop}%
\bibitem [{\citenamefont {van Leer}(1974)}]{vanLeer1974}%
  \BibitemOpen
  \bibfield  {author} {\bibinfo {author} {\bibfnamefont {B.}~\bibnamefont {van
  Leer}},\ }\bibfield  {title} {\enquote {\bibinfo {title} {{Towards the
  ultimate conservative difference scheme. II. Monotonicity and conservation
  combined in a second-order scheme}},}\ }\href {\doibase
  10.1016/0021-9991(74)90019-9} {\bibfield  {journal} {\bibinfo  {journal}
  {Journal of Computational Physics}\ }\textbf {\bibinfo {volume} {14}},\
  \bibinfo {pages} {361--370} (\bibinfo {year} {1974})}\BibitemShut {NoStop}%
\bibitem [{\citenamefont {Toro}(2009)}]{Toro1997}%
  \BibitemOpen
  \bibfield  {author} {\bibinfo {author} {\bibfnamefont {E.~F.}\ \bibnamefont
  {Toro}},\ }\href {\doibase 10.1007/b79761} {\emph {\bibinfo {title} {{Riemann
  Solvers and Numerical Methods for Fluid Dynamics}}}}\ (\bibinfo  {publisher}
  {Springer Berlin Heidelberg},\ \bibinfo {address} {Berlin, Heidelberg},\
  \bibinfo {year} {2009})\BibitemShut {NoStop}%
\bibitem [{\citenamefont {{Richard Saurel}}, \citenamefont {Petitpas},\ and\
  \citenamefont {Berry}(2009)}]{RichardSaurel2009}%
  \BibitemOpen
  \bibfield  {author} {\bibinfo {author} {\bibnamefont {{Richard Saurel}}},
  \bibinfo {author} {\bibfnamefont {F.}~\bibnamefont {Petitpas}}, \ and\
  \bibinfo {author} {\bibfnamefont {R.~A.}\ \bibnamefont {Berry}},\ }\bibfield
  {title} {\enquote {\bibinfo {title} {{Simple and efficient relaxation methods
  for interfaces separating compressible fluids, cavitating flows and shocks in
  multiphase mixtures}},}\ }\href {\doibase 10.1016/j.jcp.2008.11.002}
  {\bibfield  {journal} {\bibinfo  {journal} {Journal of Computational
  Physics}\ }\textbf {\bibinfo {volume} {228}},\ \bibinfo {pages} {1678--1712}
  (\bibinfo {year} {2009})}\BibitemShut {NoStop}%
\bibitem [{\citenamefont {Schoch}\ and\ \citenamefont
  {Nikiforakis}(2015)}]{Schoch2015a}%
  \BibitemOpen
  \bibfield  {author} {\bibinfo {author} {\bibfnamefont {S.}~\bibnamefont
  {Schoch}}\ and\ \bibinfo {author} {\bibfnamefont {N.}~\bibnamefont
  {Nikiforakis}},\ }\bibfield  {title} {\enquote {\bibinfo {title} {{Numerical
  modelling of underwater detonation of non-ideal condensed-phase
  explosives}},}\ }\href {\doibase 10.1063/1.4905337} {\bibfield  {journal}
  {\bibinfo  {journal} {Physics of Fluids}\ }\textbf {\bibinfo {volume} {27}}
  (\bibinfo {year} {2015}),\ 10.1063/1.4905337}\BibitemShut {NoStop}%
\bibitem [{\citenamefont {Souers}\ \emph {et~al.}(2009)\citenamefont {Souers},
  \citenamefont {Hernandez}, \citenamefont {Cabacungan}, \citenamefont {Garza},
  \citenamefont {Lauderbach}, \citenamefont {Liao},\ and\ \citenamefont
  {Vitello}}]{Souers2009}%
  \BibitemOpen
  \bibfield  {author} {\bibinfo {author} {\bibfnamefont {P.~C.}\ \bibnamefont
  {Souers}}, \bibinfo {author} {\bibfnamefont {A.}~\bibnamefont {Hernandez}},
  \bibinfo {author} {\bibfnamefont {C.}~\bibnamefont {Cabacungan}}, \bibinfo
  {author} {\bibfnamefont {R.}~\bibnamefont {Garza}}, \bibinfo {author}
  {\bibfnamefont {L.}~\bibnamefont {Lauderbach}}, \bibinfo {author}
  {\bibfnamefont {S.-B.}\ \bibnamefont {Liao}}, \ and\ \bibinfo {author}
  {\bibfnamefont {P.}~\bibnamefont {Vitello}},\ }\bibfield  {title} {\enquote
  {\bibinfo {title} {{Air Gaps, Size Effect, and Corner-Turning in Ambient
  LX-17}},}\ }\href {\doibase 10.1002/prep.200700232} {\bibfield  {journal}
  {\bibinfo  {journal} {Propellants, Explosives, Pyrotechnics}\ }\textbf
  {\bibinfo {volume} {34}},\ \bibinfo {pages} {32--40} (\bibinfo {year}
  {2009})}\BibitemShut {NoStop}%
\bibitem [{\citenamefont {Sheffield}\ and\ \citenamefont
  {Engelke}(2009)}]{Sheffield2009}%
  \BibitemOpen
  \bibfield  {author} {\bibinfo {author} {\bibfnamefont {S.}~\bibnamefont
  {Sheffield}}\ and\ \bibinfo {author} {\bibfnamefont {R.}~\bibnamefont
  {Engelke}},\ }\bibfield  {title} {\enquote {\bibinfo {title}
  {{Condensed-Phase Explosives: Shock Initiation and Detonation Phenomena}},}\
  }in\ \href {\doibase 10.1007/978-3-540-77080-0_1} {\emph {\bibinfo
  {booktitle} {Shock Wave Science and Technology Reference Library, Vol. 3}}}\
  (\bibinfo  {publisher} {Springer Berlin Heidelberg},\ \bibinfo {address}
  {Berlin, Heidelberg},\ \bibinfo {year} {2009})\ pp.\ \bibinfo {pages}
  {1--64}\BibitemShut {NoStop}%
\bibitem [{\citenamefont {Tarver}\ and\ \citenamefont
  {McGuire}(2002)}]{Tarver2002a}%
  \BibitemOpen
  \bibfield  {author} {\bibinfo {author} {\bibfnamefont {C.~M.}\ \bibnamefont
  {Tarver}}\ and\ \bibinfo {author} {\bibfnamefont {E.~M.}\ \bibnamefont
  {McGuire}},\ }\bibfield  {title} {\enquote {\bibinfo {title} {{Reactive flow
  modeling of the interaction of TATB detonation waves with inert
  materials}},}\ }\href@noop {} {\bibfield  {journal} {\bibinfo  {journal}
  {Twelfth International Detonation}\ ,\ \bibinfo {pages} {1--10}} (\bibinfo
  {year} {2002})}\BibitemShut {NoStop}%
\end{thebibliography}%

\end{document}